\documentclass[a4paper,10pt, twocolumn]{revtex4-2}
\usepackage[utf8]{inputenc}
\usepackage{graphicx}
\usepackage[hidelinks]{hyperref}
\usepackage{amsmath}
\usepackage{amssymb}
\usepackage{xcolor}
\usepackage{xspace}

\hypersetup{
    colorlinks,
    linkcolor={red!50!black},
    citecolor={blue!50!black},
    urlcolor={blue!80!black}
}

\newcommand{\ff}[2]{f^T_{#1 \rightarrow #2}}
\newcommand{\hff}[2]{\hat f^T_{#1 \rightarrow #2}}

\newcommand{\hfpf}[2]{\hat f'^T_{#1 \rightarrow #2}}

\newcommand{\fb}[2]{f^T_{#2 \rightarrow #1}}

\newcommand{\NFRs}{NFRs\xspace}
\newcommand{\NFR}{NFR\xspace}

\newcommand{\rr}{\tilde{\rho}}

\newcommand{\Cij}{\C{i}{j}}
\newcommand{\C}[2]{\Delta_{#1 \rightarrow #2}}
\newcommand{\Cbi}[2]{\Delta_{#1 \Rightarrow #2}}
\newcommand{\Ctri}[3]{\Delta_{#1 \underset{#3}{\rightarrow} #2}}
\newcommand{\Cshort}[3]{\Delta_{#1 \rightarrow #2}}
\newcommand{\Cp}[2]{\Delta'_{#1 \rightarrow #2}}
\newcommand{\Cbip}[2]{\Delta'_{#1 \Rightarrow #2}}
\newcommand{\Ctrip}[3]{\Delta'_{#1 \underset{#3}{\rightarrow} #2}}
\newcommand{\R}[2]{R_{#1 \rightarrow #2}}

\newcommand{\nuz}{\hat\nu^0}
\renewcommand{\u}{u}

\newcommand{\ec}{\epsilon_{\textrm{\tiny C}}}
\newcommand{\eo}{\epsilon_{\textrm{\tiny OO}}}
\newcommand{\er}{\epsilon_{\textrm{\tiny RF}}}
\newcommand{\eb}{\epsilon_{\textrm{\tiny b}}}

\newcommand{\frep}{f_\textrm{\tiny rep}}
\newcommand{\hfrep}{\hat f_\textrm{\tiny rep}}

\newcommand{\RF}{\textrm{RF}}
\newcommand{\RFt}{\textrm{\tiny RF}}
\newcommand{\OO}{\textrm{OO}}
\newcommand{\OOt}{\textrm{\tiny OO}}
\newcommand{\OC}{\textrm{OC}}
\newcommand{\OCt}{\textrm{\tiny OC}}
\newcommand{\MC}{\textrm{MW}}
\newcommand{\MCt}{\textrm{\tiny MW}}
\newcommand{\CL}{\textrm{CL}}
\newcommand{\CLt}{\textrm{\tiny CL}}
\newcommand{\clock}{{\textrm{\tiny clock}}}

\newcommand{\ex}[1]{\times10^{#1}}

\begin{document}

\title{Universal formalism for data sharing and processing in clock comparison networks}

\author{Jérôme Lodewyck}
\email[]{jerome.lodewyck@obspm.fr}
\affiliation{LNE–SYRTE, Observatoire de Paris, Université PSL, CNRS, Sorbonne Université, 61 avenue de l'Observatoire, F-75014 Paris, France}
\author{Erik Benkler}
\affiliation{Physikalisch-Technische Bundesanstalt, Bundesallee 100, D-38116 Braunschweig, Germany}
\author{Jochen Kronjäger}
\affiliation{National Physical Laboratory, Hampton Road, Teddington, Middlesex, TW11 0LW, United Kingdom}
\author{Sebastian Koke}
\affiliation{Physikalisch-Technische Bundesanstalt, Bundesallee 100, D-38116 Braunschweig, Germany}
\author{Rodolphe Le Targat}
\affiliation{LNE–SYRTE, Observatoire de Paris, Université PSL, CNRS, Sorbonne Université, 61 avenue de l'Observatoire, F-75014 Paris, France}
\author{Paul-Eric Pottie}
\affiliation{LNE–SYRTE, Observatoire de Paris, Université PSL, CNRS, Sorbonne Université, 61 avenue de l'Observatoire, F-75014 Paris, France}

\begin{abstract}
	We present a mathematical framework adapted to the comparison of atomic clocks remotely connected by a complex network of optical or microwave links. This framework facilitates the computation of frequency ratios using a generic set of equations that are valid for a wide variety of clock architectures, making the comparison of a large number of clocks practical. This formalism suggests a generic data exchange protocol that can be used in clock comparison collaborations, for which data logging and decision taking are completely local procedures, independently implemented by the participants.
\end{abstract}

\maketitle

\section{Introduction}

\begin{figure*}
	\includegraphics[width=0.9\textwidth]{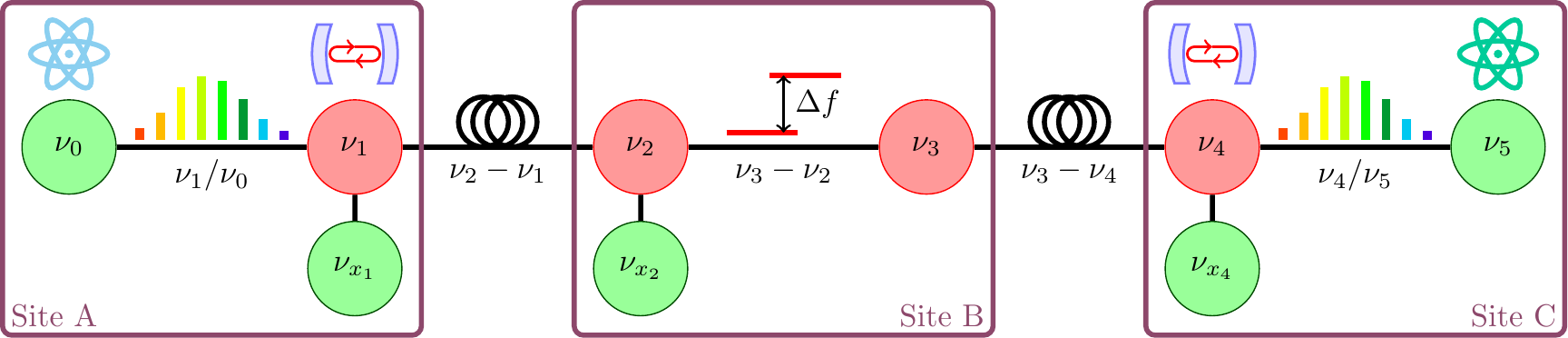}
	\caption{\label{fig:europeannetwork}Typical branch of the elementary network connecting European metrology institutes. The measurement chain consists of the two remote clocks at the end points A and C, with frequency $\nu_0$ and $\nu_5$, respectively. Each clock frequency is compared to the frequency ($\nu_1$ and $\nu_4$, respectively) of a cavity stabilised ultra-stable laser source, via an optical frequency comb.
    These laser sources, typically operating around $194.4$~THz, near the minimum of fiber attenuation, 	are transmitted by phase-stabilised optical fiber links to a remote site B, for example a data center, where they arrive with frequencies $\nu_2$ and $\nu_3$ respectively. The phase-stabilising mechanism typically induces fixed additive frequency shifts $\nu_2 - \nu_1$ and $\nu_3 - \nu_4$, which are derived from RF references local to the endpoints A and C (here labeled $\nu_{x_1}$ and $\nu_{x_4}$). At the remote site B, the frequency difference $\Delta f = \nu_3 - \nu_2$ is measured with respect to a local RF reference, whose frequency $\nu_{x_2}$ is steered by a GPS receiver. In this network, we represent the link lasers by red nodes because their frequencies cannot usually be accurately anticipated, either because they are referenced to arbitrary and fluctuating artefacts (ultra-stable cavities), or because they are arbitrarily shifted over the span of the network. In contrast, the clocks and RF references are represented in green to reflect their link to accurate references. This notion will be rigorously defined and exploited in the framework proposed in this paper.
    The frequency ratio between the compared clocks (here $\nu_5/\nu_0$) is derived by combining the successive intermediate measurements along the network branch.
	In a full network, several of these branches must be considered: for instance, in a network comprising 10 clocks, as many as 45 branches must be calculated.
}
\end{figure*}

Worldwide time and frequency comparison of clocks has a long-standing history and is the backbone for steering the International Atomic Time (TAI) by primary and secondary frequency standards using the local implementations UTC(k) as a pivot~\cite{Circular_T,bau06}. While the uncertainty of the established microwave satellite connections is sufficient to compare the best microwave clocks realizing the SI second~\cite{Fujieda_2018}, they fail in comparing frequencies of optical clocks at the level of their uncertainty~\cite{dro15}. Over the last decades phase-stabilised optical fiber links~\cite{ma94, wil08a, chi15, rau15, cal14a} have been investigated and their capability for remote optical clock frequency comparisons with negligible statistical and systematic uncertainty contribution from the link has been demonstrated both at metropolitan scale~\cite{Yamaguchi_2011} and continental scale~\cite{lisdat2016clock}. Furthermore, the averaging time required to reach the regime where instability is dominated by the clocks can be orders of magnitude shorter with phase-stabilised optical fiber links instead of microwave satellite links~\cite{Fujieda_2018, rie20}. In a joint effort, national metrology institutes in Europe have established an elementary network of fiber links~\cite{chi15, kok19, Kronjaeger2016} connecting their sites in Braunschweig, Paris and London. This network, schematically represented in figure~\ref{fig:europeannetwork}, now enables the simultaneous frequency comparison of many optical and microwave clocks~\cite{lisdat2016clock, gue17,del17}.

Performing regular frequency comparisons of both optical and microwave clocks in such a network requires an appropriate formalism that is on the one hand accurate and on the other hand flexible enough to cope with the different infrastructure elements and clock types, and eventually can incorporate microwave links as well.
While it is feasible to work out \emph{ad hoc} equations to calculate the frequency ratio between two remote optical clocks connected by a single optical fiber link from the output of the experimental apparatuses involved in the comparison, the situation becomes rapidly more complex when a large number of clocks are connected by different links with different architectures, operated by a diverse collaboration, and with multiple intermediate steps where data is collected by various parties. For these comparisons, a more rigorous and generic approach must be followed in order to facilitate the data analysis.

In this paper, we report on and discuss the foundations of a formalism useful for performing regular clock comparisons in such collaborations. It can be used to compute the value of the frequency ratios with high precision between any pair of oscillators in the network, under the sole assumption that an \emph{a priori} approximate value for these frequency ratios is available. This formalism has several features adapted to large clock networks. First, it is generic: it can accommodate an arbitrary network architecture, in which any optical or microwave clocks can be compared through phase-coherent RF or optical links. Links may consist of a single phase-stabilised transmission line or involve intermediate nodes where phase or frequency measurements are recorded.
Second, it is precise: although the formalism yields approximate values of frequency ratios, the approximation error is negligible when considering the current or foreseeable uncertainty of optical clocks. Furthermore, data can be shared and combined using only double precision floating point arithmetics.
Finally, it is practical: it defines a generic data exchange format that requires the production of a single and local quantity per measurement device within the network. By sharing these data, the frequency ratios between any pair of clocks or oscillators in the network can be calculated using equations that take into account the topology of the network, but are independent of the details of the implementation. Thus, such a well defined protocol facilitates the data evaluation and helps reducing and correcting mistakes.

The paper is organized as follows: sections~\ref{sec:network} and~\ref{sec:networkcomponents} introduce the concepts and notations used in this paper, and relate them to the actual physical implementation of the network components. In section~\ref{sec:computingratios}, we derive the fundamental equations used to compute frequency ratios between the oscillators in the clock network. In sections~\ref{sec:oscillatorcentric} and~\ref{sec:TBF}, we cast these equations in a configuration focused on the oscillators composing the network, or using a terminology based on transfer beats, respectively. Finally, section~\ref{sec:comparing} discusses the approximation error introduced by the formalism. A series of appendices proposes additional interpretations of the formalism. In appendix~\ref{sec:usecase}, we illustrate the equations derived in this paper by applying them to a fictitious network.

\section{Network architectures}
\label{sec:network}

\begin{figure}
	\begin{center}
		\includegraphics[width=\columnwidth]{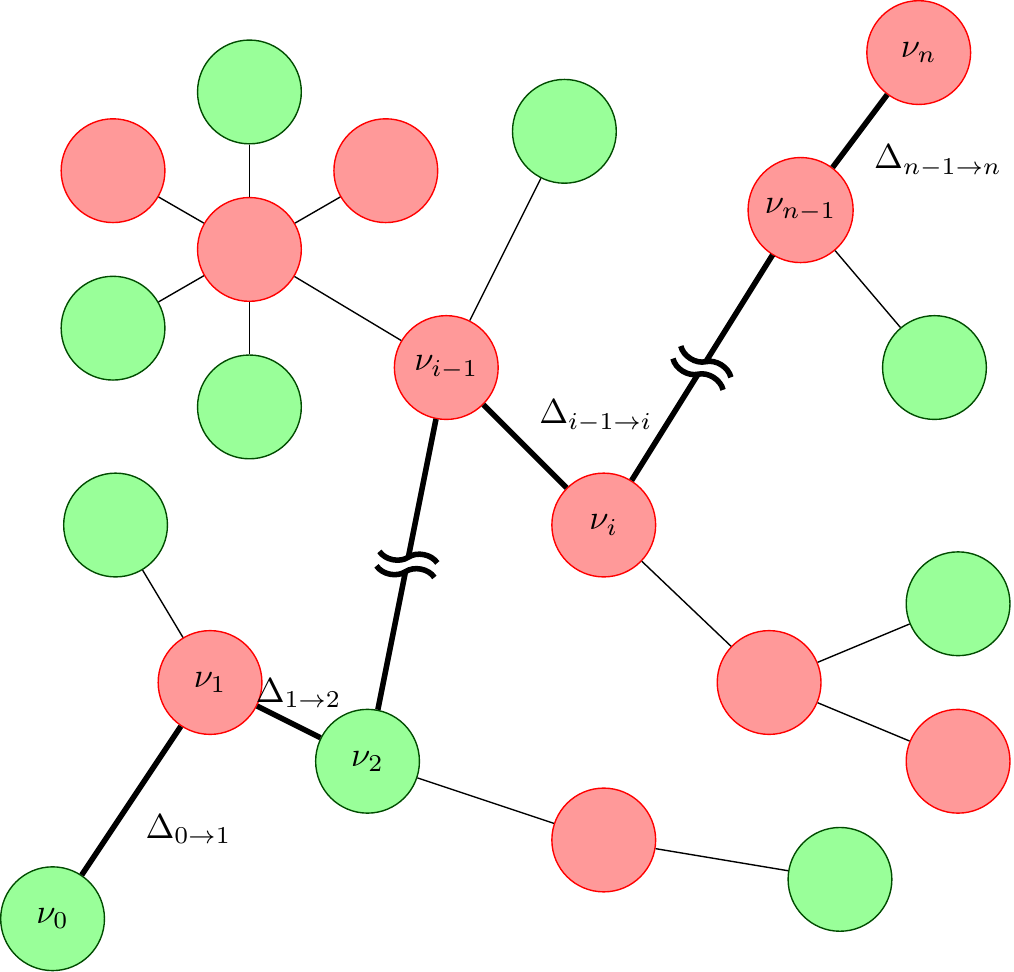}
	\end{center}
	\caption{\label{fig:network}
	Generic network model allowing to describe the setup in figure~\ref{fig:europeannetwork}, along with other types of clocks and links, such as microwave clocks or links. Round nodes denote microwave or optical oscillators. These oscillators are linked by comparators. The comparator connecting the oscillators $i$ and $j$ yields a numerical outcome $\Cij$. We highlight a specific path through the network (thick lines) linking oscillator $0$ to oscillator $n$ with respective frequencies $\nu_0$ and $\nu_n$ via a series of $n$ independent comparisons with outcomes $\C{0}{1}$ to $\C{n-1}{n}$. These outcomes are then used to compute the frequency ratio $\nu_n/\nu_0$.  This computation will rely on the assumption that there is a set of oscillators, represented in green, whose frequency ratios are precisely known in advance (typically down to $10^{-13}$), with methods external to the core operation of the network (\emph{e.g.} through the realization of a unit, or by GPS disciplining). These oscillators are said ``accurate'' because they are the generalization to the generic network of the accurate references introduced in figure~\ref{fig:europeannetwork}.}
\end{figure}

The generic network model underlying the formalism is depicted in figure~\ref{fig:network}. It is an abstract generalization of the network represented in figure~\ref{fig:europeannetwork}, composed of oscillators having a frequency $\nu_i$, connected by comparators. Oscillators can be clocks or dedicated oscillators such as cavity-stabilised lasers, or may simply represent the frequency of the signal at a specific point of the network. Comparators acquire information about the relation of a specific pair of oscillators in the network based on local measurement outcomes.

Physical implementations of comparators may for example consist of frequency combs measuring an optical frequency ratio, actively stabilised fiber links with a fixed frequency offset or frequency counters recording a beat frequency. Examples of implementations are discussed in detail in section~\ref{sec:comparatorexamples}. Regardless of implementation, the measurement outcome can be described by one or more frequency ratios, with frequency measurements, such as those produced by counters, being a special case of ratios involving a local RF reference.

The comparator linking the oscillators $i$ and $j$ outputs a stream of data noted $\C{i}{j}$. We assume the measurement data at all comparators to be simultaneously sampled~\footnote{The impact of non-simultaneous measurements has been addressed in the supplement of~\cite{lisdat2016clock}} and omit the explicit time dependence in all equations. Our goal in this paper is to find a suitable definition for $\C{i}{j}$, and to propose generic equations to calculate remote frequency ratios between clocks and oscillators from these quantities. In order to have a generic scheme for publishing comparator data, $\C{i}{j}$ should be local: it may depend on the frequencies $\nu_i$ and $\nu_j$ of the oscillators it compares and on the frequency of a local RF reference, but not on the frequency of any other oscillator in the network, nor on the specific path under consideration in the network. The full set of local outcomes $\C{i}{j}$ should be sufficient to compute the frequency ratio between any pair of oscillators in the network.

When comparing the oscillators composing the network, the fundamental quantities we are interested in are frequency ratios. We define:
\begin{equation}
	\label{eq:rho}
	\rho_{i,j} \equiv \frac{\nu_i}{\nu_j}
\end{equation}
the frequency ratio between the oscillators $i$ and $j$. Using the frequency ratios $\rho_{i,j}$, a simple way to compare the oscillator $n$ with the oscillator $0$ is to write:
\begin{equation}
	\label{eq:rhon0}
	\rho_{n,0} = \prod_{i=1}^n \rho_{i, i-1},
\end{equation}
where $\rho_{i, i-1}$ is the frequency ratio between the oscillators $i$ and $i-1$ measured locally by the comparator linking $i-1$ to $i$. The fractional uncertainty of the frequency ratio $\rho_{n,0}$ is thus given by the combined fractional uncertainty of the individual frequency ratios in the product.

However, the direct implementation of equation~(\ref{eq:rhon0}) suffers from two problems. First, it requires manipulating numbers with a numerical precision higher than the clock uncertainty. The ubiquitous IEEE 754 double precision floating point representation, which is the current standard for floating point arithmetic in numerical-analysis software and hardware, has a $10^{-16}$ fractional rounding error and is therefore not sufficient to  compare optical clocks with their current uncertainty of $10^{-18}$, thus requiring the use of specialised arbitrary precision computing libraries. The second and more stringent issue is that frequency ratios are generally not locally accessible where optical oscillators are interfaced without a frequency comb, for example when measuring a frequency beatnote between two lasers, or when applying a frequency offset with an acousto-optic modulator, with respect to a local RF reference (see figure~\ref{fig:europeannetwork}). Therefore, in practice, equation~(\ref{eq:rhon0}) cannot be used as it is.

The way out of this problem is to notice that if, at a particular node the frequency difference $\Delta f = \nu_{i} - \nu_{i-1}$ is induced or measured between oscillators $i$ and $i-1$, then the frequency ratio $\rho_{i, i-1}$, part of the product~(\ref{eq:rhon0}), reads:
\begin{equation}
	\label{eq:rhon0Df}
	\rho_{i,i-1} = 1+\frac{\Delta f}{\nu_{i-1}}.
\end{equation}
Given that $\Delta f/\nu_{i-1} \ll 1$, typically on the order of $10^{-6}$ or below, a relative uncertainty of $10^{-13}$ on $\Delta f$ and $\nu_{i-1}$ is sufficient to reach a relative uncertainty of $10^{-19}$ on $\rho_{n,0}$. The frequency difference $\Delta f$ can thus be referenced to GPS or to an H-Maser, while $\nu_{i-1}$ can be traced back to the frequency $\nu_0$ of oscillator $0$ by calculating $\rho_{i-1,0}$ first, where an estimate of $\nu_0$ with $10^{-13}$ fractional uncertainty must be available.

Although this procedure is valid, it does not fulfill our initial goal to write generic equations for arbitrary network architectures, because it requires a specific treatment for some comparators.

In order to establish a generic formalism, we can represent the frequency ratio $\rho_{i,j}$ by a small relative deviation, noted $\rr_{i,j}$ and called ``Reduced Frequency Ratio'' (RFR) in this paper, from a constant \emph{a priori} estimate $\rho^0_{i,j}$ of $\rho_{i,j}$, here called ``Nominal Frequency Ratio'' (NFR):
\begin{equation}
	\label{eq:rhotorr}
	\rho_{i,j} = \rho^0_{i,j}(1+\rr_{i,j}).
\end{equation}
This representation is routinely employed for the comparisons of clocks by satellites, because it effectively solves the numerical issue of representing numbers with many significant digits, and abstracts the frequency scaling of the RF local oscillators used as a pivot. For instance, this notation is used to publish TAI calibrations by primary and secondary frequency standards in the BIPM (Bureau International des Poids et Mesures) circular T~\cite{PETIT2015480}~; Reference~\cite{rie20} is a recent example of the usage of RFRs in the context of optical clocks comparisons by satellites. The numerical handling is further simplified because replacing the frequency ratios that appear in equation~(\ref{eq:rhon0}) by their expression~(\ref{eq:rhotorr}) and expanding their product, yields the remote frequency ratio $\rho_{n,0}$ as a linear combination of local RFRs, provided that products of RFRs are negligible. This simplification is valid for the comparison of clocks by satellite techniques, for which NFRs marginally deviate from the actual frequency ratios. However, it generally fails for the remote comparison of clocks through optical fiber links because the optical local oscillators transmitted in fiber links (\emph{i.e.} cavity-stabilized lasers) usually exhibit large frequency offsets and drifts along a network path, as represented in figure~\ref{fig:europeannetwork}. In the next sections, we will show that this linear expression nevertheless holds for specific network layouts in which known frequency references are available.

Introducing RFRs also resolves the problem raised above that fiber networks incorporate heterogeneous comparison methods, some yielding frequency ratios, other yielding frequency offsets. Indeed, the RFR $\rr_{i,j}$ can be written as:
\begin{equation}
	\rr_{i,j} = \frac{\nu_i - \rho^0_{i,j}\nu_j}{\rho^0_{i,j}\nu_j}.
\end{equation}
The numerator of this expression is the so-called ``transfer-beat'' that is often used with frequency combs~\cite{ste02a, tel02b}, and that generalizes the simple optical beatnote $\Delta f$. Hence, RFRs can represent in a unified way devices that naturally measure frequency ratios, and devices that produce transfer beats, including frequency offsets. Here again, the network layout must include known local references against which the transfer beats are measured.

In the next section, we will present a rigorous foundation for the concepts sketched above, starting with a definition of the network layout and components that are covered by the formalism. The local frequency references play a central role in this architecture, and we will discuss how the error of these references impacts the computation of remote frequency ratios (section~\ref{sec:comparing}).

The remote frequency ratios will be written as linear combinations of transfer beats (section~\ref{sec:TBF}) reminiscent of the notation used for frequency combs, but also, in a particular case, as a simple sum of RFRs (section~\ref{sec:oscillatorcentric}). This dual notation thus reconciles these two points of view.

\section{Network components}
\label{sec:networkcomponents}

In this section, we introduce the notation that we will use in this document, and model the components of the network. These components are first introduced as a formal abstraction (sections~\ref{sec:nrfr} and~\ref{sec:comparatormodels}), and later associated with actual experimental apparatuses (section~\ref{sec:comparatorexamples}). The reader may also refer to figures~\ref{fig:europeannetwork} and~\ref{fig:usecase}, which propose examples of network implementations.

\subsection{Nominal and reduced frequency ratios}
\label{sec:nrfr}

The formalism developed in this paper relies on the fundamental assumption that \emph{a priori} estimates $\rho^0_{i,j}$ of frequency ratios $\rho_{i,j}$ between selected oscillators $i, j$ of the network are available. The quantities $\rho^0_{i,j}$, hereafter called the ``Nominal Frequency Ratios'' (\NFRs), are exactly known numerical constants. We assume that the \NFRs are chosen such that
\begin{equation}
    \rho^0_{i,j}\,\rho^0_{j,i} = 1,
\end{equation}
which is physically equivalent to the path invertibility of the comparator, and in particular includes $\rho^0_{i,i}= 1$.

Being dimension-less numbers, the \NFRs are not \emph{a priori} related to any frequency unit. They can rather be arbitrarily chosen by the operators of the network, as long as they are close enough to the actual frequency ratios $\rho_{i,j}$. More specifically, the formalism presented in this paper requires that two sets of \NFRs are defined:
\begin{enumerate}
	\item A comparator links two oscillators $i$ and $j$, and yields an output $\C{i}{j}$, possibly with the help of a third oscillator, as detailed in section~\ref{sec:comparatormodels}. The NFRs between these three oscillators (the ``local \NFRs'') must be defined.
	They are used by the comparator as \emph{a priori} knowledge about the frequency ratios between the oscillators in order to compute $\C{i}{j}$. This local computation should be realizable using only the local \NFRs. Moreover, these local \NFRs are, in the most general case, independent of other local \NFRs, meaning that the operator of the comparator can choose their values at will, without being prejudiced by the choice of the \NFRs made for other comparators.
	\item There is a subset of oscillators for which the \NFRs between any two of these oscillators are defined. We require that all these \NFRs are particularly close to their corresponding actual frequency ratios. To reflect this, we call these oscillators ``accurate'', in the sense that the \NFRs constitute a very good anticipation of frequency ratios. These oscillators will act as references across the network. These NFRs are used when combining the outputs of several comparators to calculate the frequency ratio between remote oscillators.
\end{enumerate}
Using the \NFRs, we define the reduced frequency ratios (RFRs) as:
\begin{equation}
	\label{eq:rr}
 	\rr_{i,j} \equiv\frac{\rho_{i,j}}{\rho^0_{i,j}} - 1,
\end{equation}
The RFR $\rr_{i,j}$ describes how much the actual frequency ratio $\rho_{i,j}$ deviates from the \NFR $\rho^0_{i,j}$. Because $\rho^0_{i,j}$ must be close to $\rho_{i,j}$, we have:
\begin{equation}
	\label{eq:rrll1}
	\quad |\rr_{i,j}| \ll 1, \quad \textrm{and specifically} \quad  \rr_{i,i} = 0.
\end{equation}
In the practical examples below, we will have $|\rr_{i,j}| \lesssim 10^{-6}$ for all oscillators, and $|\rr_{i,j}| \lesssim 10^{-13}$ for all accurate oscillators. With these bounds, clocks and RF local references are accurate oscillators because their frequency ratios can easily be estimated to below $10^{-13}$, while cavity-stabilized optical oscillators are inaccurate because their daily relative frequency drift is usually orders of magnitude larger than $10^{-13}$. These bounds are only example orders of magnitude, that can be adapted to actual networks depending on the target uncertainty of the computed frequency ratios, as detailed in section~\ref{sec:comparing}.

Finally, the RFRs can be handled with a reduced number of significant digits because the most significant digits of the frequency ratios are removed.

\subsection{Comparator model}
\label{sec:comparatormodels}

A comparator is a device linking two oscillators $i$ and $j$, which is able to output a numerical value $\C{i}{j}$ that is useful to compute frequency ratios over the network by combining the various comparator outputs. As outlined in section~\ref{sec:network}, a comparator may, for example, be a device that directly measures a frequency ratio, or a device that measures beatnotes between optical oscillators, referenced to an RF local oscillator. In order to encompass all possible cases, we formally define a generic comparator as a tripartite system comprising the two oscillators $i$ and $j$, as well as a third,  auxiliary oscillator $x$ that is used as a reference. We define the comparator output as the differential relative measurement of $i$ and $j$ with respect to $x$, reading:
\begin{equation}
	\label{eq:Cij3lin}
	\Ctri{i}{j}{x} \equiv \rr_{j,x} - \rr_{i,x} \qquad \textrm{where $x$ is accurate.}
\end{equation}
In this definition, we require that the reference oscillator $x$ is accurate, in the way defined in section~\ref{sec:nrfr}, because this clause will prove to be necessary in order to use the comparator outputs to compute remote frequency ratios within the formalism presented in this paper (section~\ref{sec:computingratios}).

Generating the comparator output requires that the NFRs $\rho^0_{i,x}$ and $\rho^0_{j,x}$ are defined, as presented in section~\ref{sec:nrfr}. However, the comparator's measurement capability may or may not allow to determine $\rr_{j,x}$ and $\rr_{i,x}$ separately. Examples of how the comparator output can be determined from measurement outcomes will be given in section~\ref{sec:comparatorexamples} below.

\begin{figure}
	\begin{center}
		\includegraphics[width=\columnwidth]{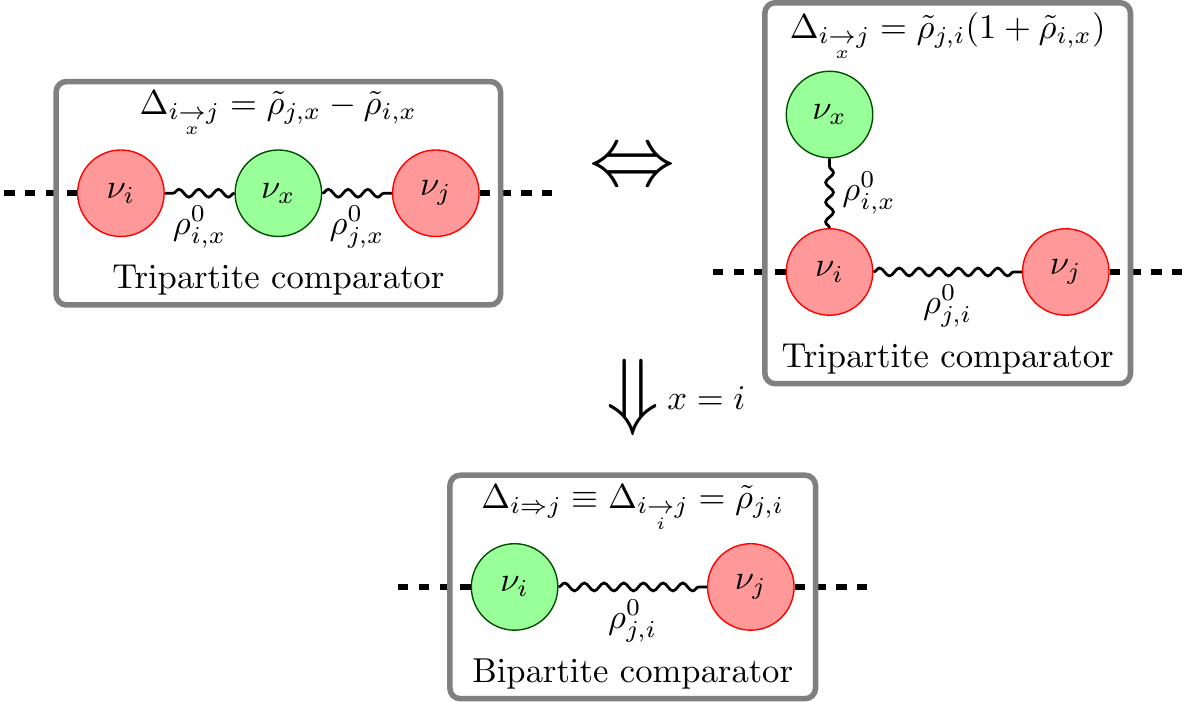}
	\end{center}
	\caption{\label{fig:comparatormodels}Sketch of comparators comparing the oscillators $i$ and $j$. A tripartite comparator makes use of an auxiliary accurate oscillator $x$, that can equivalently be connected to one of the two oscillators or inserted between the two oscillators. A bipartite comparator is a specific case of a tripartite comparator in which the accurate oscillator $x$ is identified with one of the two other oscillators. The comparator output $\Ctri{i}{j}{x} = -\Ctri{j}{i}{x}$ can be calculated from RFRs between the oscillators. The black waves show the NFRs that must be defined for these RFRs to be well defined}
\end{figure}

Like the RFRs, the comparator output is a small quantity:
\begin{equation}
	|\Ctri{i}{j}{x}| \ll 1,
\end{equation}
and the forward and backward outputs are related by:
\begin{equation}
	\Ctri{j}{i}{x} = -\Ctri{i}{j}{x}.
\end{equation}
Using the definition of the RFRs, we can derive from the definition~(\ref{eq:Cij3lin}) alternative expressions for the comparator output:
\begin{equation}
	\label{eq:CijT}
	\Ctri{i}{j}{x} = \rr_{j,i}(1+\rr_{i,x}) = -\rr_{i,j}(1+\rr_{j,x}),
\end{equation}
where the NFR $\rho^0_{j,i}$ that underpins the RFR $\rr_{j,i}$ is defined by:
\begin{equation}
	\rho^0_{j,i} \equiv \rho^0_{j,x}\rho^0_{x,i}.
\end{equation}
NFRs obeying this chain rule are called ``transitive''. The three local NFRs are chosen transitive, so that they are all three unambiguously defined within the local set. In the following, we call $\rho^0_{j,i}$ the ``linkage NFR'' of the comparator linking the network nodes $i$ and $j$, in order to distinguish it from the other two local NFRs $\rho^0_{j,x}$ and $\rho^0_{i,x}$ linking the network nodes to the auxiliary accurate oscillator $x$. Equation~(\ref{eq:CijT}) shows that the comparator output is the reduced frequency ratio $\rr_{j,i}$, slightly corrected by a factor that relates the two oscillators $i$ and $j$ to the auxiliary oscillator $x$, as depicted in figure~\ref{fig:comparatormodels}. It follows that, when replacing the reference oscillator $x$ with another reference $x'$, the comparator output is altered by:
\begin{equation}
	\label{eq:deltaDelta}
	\Ctri{i}{j}{x'} - \Ctri{i}{j}{x} \simeq \rr_{j,i}\rr_{x,x'}.
\end{equation}
 With the orders of magnitude given above for the RFRs, and considering that $x$ and $x'$ are accurate oscillators, the difference $|\Ctri{i}{j}{x'} - \Ctri{i}{j}{x}| \lesssim 10^{-19}$ is negligible. Therefore, the comparator output is mostly independent of $x$, and we will simply note it $\C{i}{j}$, unless we want to specifically mention the reference oscillator.

When the accurate reference $x$ is one of the compared oscillators itself, we say that the comparator is bipartite. The output of a bipartite comparator is obtained by identifying $x$ with $i$ or $j$ in the definition~(\ref{eq:Cij3lin}):
\begin{eqnarray}
	\label{eq:Cij2}
	\Cbi{i}{j} \equiv \Ctri{i}{j}{i} &=& \rr_{j,i}  \quad \textrm{if $i$ is accurate.}
\end{eqnarray}
Note the difference in notation between bipartite $\Cbi{i}{j}$ and tripartite $\Ctri{i}{j}{x}$, and that $\Cbi{i}{j} \neq -\Cbi{j}{i}$, given that $\Ctri{i}{j}{i} = -\Ctri{j}{i}{i} \neq -\Ctri{j}{i}{j}$, the reference being different. Consequently, any bipartite comparator is capable of directly measuring the frequency ratio $\rho_{j,i}$ between an oscillator and an accurate oscillator.

Any bipartite comparator can be represented as a tripartite one, but not vice-versa. However, the relations derived above show that the output of a tripartite comparator can in fact be written as the sum of two bipartite comparator outputs if measurement capability allows to determine $\rr_{j,x}$ and $\rr_{i,x}$ separately:
\begin{equation}
	\label{eq:CisCC}
	\Ctri{i}{j}{x} = \Cbi{x}{j} - \Cbi{x}{i}.
\end{equation}

In this case, a tripartite comparator can be reduced to the combination of two cascaded bipartite comparators. The operator of such a tripartite comparator may equivalently publish $\Ctri{i}{j}{x}$ as a whole, or publish both $\Cbi{x}{i}$ and $\Cbi{x}{j}$. This possibility will be exploited in appendix~\ref{sec:exactfr}.

When $\Cbi{x}{i}$ and $\Cbi{x}{j}$ cannot be measured independently, we say that the comparator is genuinely tripartite. In this case, the reference oscillators $x$ is not connected by a chain of comparator outputs to the other oscillators of the network: the only knowledge about the frequency of $x$ available to the network is contained in the NFRs between accurate oscillators, which is chosen such that the RFRs between these accurate oscillators is upper bounded. This upper bound can be guaranteed by \emph{e.g.} steering the frequency of the local reference $x$ to a GPS signal.

In order to treat tripartite and bipartite comparators on equal footing, we have required that one of the oscillators -- the auxiliary oscillator in the tripartite case or one of the oscillators in the bipartite case -- is accurate. This trade-off is required in order to homogeneously describe comparators capable of determining a ratio and the ones inducing or measuring a frequency difference. We will later discuss strategies to mitigate this limitation. In particular, we propose in appendix~\ref{sec:exactfr} an extension of the formalism which can incorporate reference-free bipartite comparators, \emph{i.e.} comparators directly measuring the frequency ratio between two  oscillators without the requirement that one of them must be accurate.

\begin{figure*}
	\begin{center}
		\includegraphics[width=0.8\textwidth]{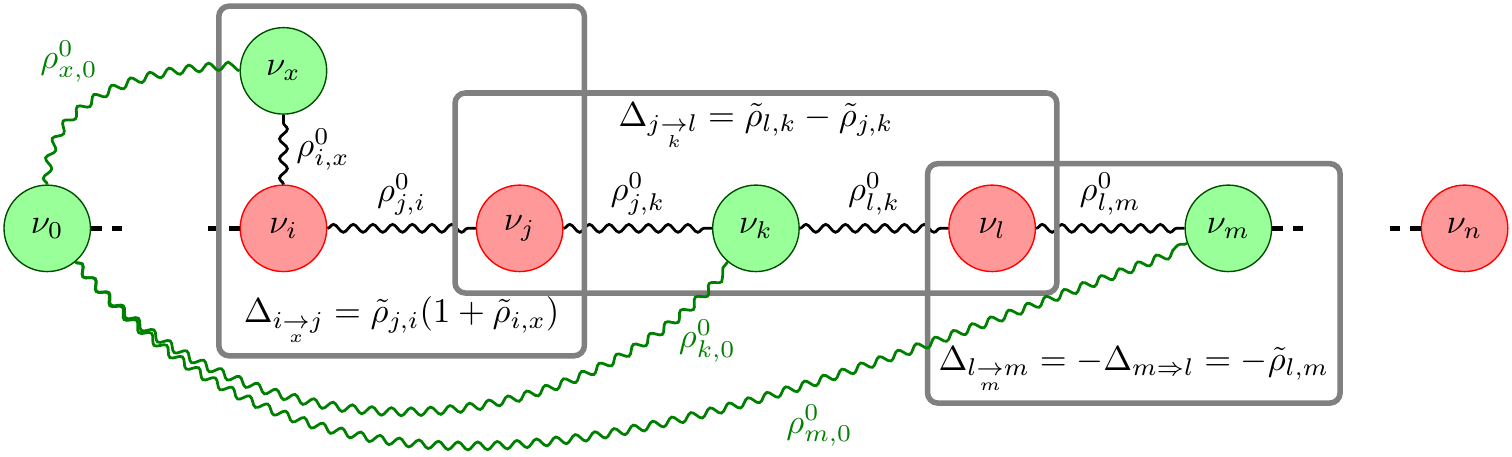}
	\end{center}
	\caption{\label{fig:networkcomponents}Assembling the network components. In order to compare the oscillator $n$ to the accurate oscillator $0$, the formalism developed here requires the definition of NFRs, represented by waves in this figure. NFRs represented in black, following figure~\ref{fig:comparatormodels}, are internal to comparators, and used to produce the comparator output. We represent in green the NFRs between accurate oscillators, which are  used to assemble the comparator outputs into remote frequency ratios. In the example drawn here, three different types of comparators are connected. They all include an accurate oscillator, as required by the formalism. Note that at the network level, the defined NFRs form loops, \emph{i.e.} it is possible to travel from one oscillator to the other following different paths along which NFRs are defined. This topological property will be relevant in section~\ref{sec:oscillatorcentric} and appendix~\ref{sec:exactfr}.}
\end{figure*}

The various components we have defined are depicted in figure~\ref{fig:networkcomponents}. Having defined the comparator output, we now have to ensure that this output can be actually produced from local measurements in practical networks (section~\ref{sec:comparatorexamples}), and that frequency ratios between oscillators in the network can be deduced from a combination of these outputs (section~\ref{sec:computingratios}).

\subsection{Examples of comparators}
\label{sec:comparatorexamples}

\paragraph{Comparing optical oscillators using a frequency beatnote.} We consider the comparison between two optical oscillators $\OO_1$ and $\OO_2$, \emph{e.g.} cavity stabilized ultra-stable infrared lasers, whose frequencies are close enough such that the beatnote between the two lasers is counted by a frequency counter referenced to an RF local oscillator, forming a genuinely tripartite comparator with the RF reference taking the role of the auxiliary accurate oscillator $x$. Choosing the NFRs such that
\begin{equation}
	\rho^0_{\OOt_1,\RFt} = \rho^0_{\OOt_2,\RFt} \quad \Leftrightarrow \quad \rho^0_{\OOt_1,\OOt_2} = 1,
\end{equation}
the comparator output reads, according to the relation~(\ref{eq:Cij3lin}):
\begin{equation}
	\Ctri{\OOt_1}{\OOt_2}{\RFt} = \rr_{\OOt_2,\RFt} - \rr_{\OOt_1,\RFt} = \frac{\nu_{\OOt_2} - \nu_{\OOt_1}}{\nu_\RFt}\rho^0_{\RFt,\OOt_1}.
\end{equation}
The first factor in this comparator output is experimentally accessible from the measurement of the optical beatnote with the frequency counter referenced to the RF reference, and the NFR $\rho^0_{\RFt,\OOt_1}$ is a known constant.

A comparator consisting of a constant frequency shift generated by \emph{e.g.} a frequency synthesizer referenced against the RF reference and applied to the laser beam is described in the same way. In this case, the first factor of the comparator output is the multiplication ratio of the frequency synthesizer.

This comparator has two prominent features. First, it is a genuinely tripartite comparator that cannot be reduced to a succession of two bipartite comparators, because the hardware used in this comparator does not yield both $\Cbi{x}{\OOt_1}$ and $\Cbi{x}{\OOt_2}$ independently. This means in particular that the RF reference oscillator is not connected  to other oscillators of the network. Second, we have set a constraint on the NFRs, such that they cannot be arbitrarily chosen. This constraint defines the  magnitude of the  comparator output:~for the comparator output to be kept below $10^{-6}$, the beatnote frequency must not exceed a few 100~MHz, because $\rho^0_{\RFt,\OOt_1}/\nu_\RFt$ is on the order of an inverse optical frequency.

Both features are unique among all the examples of comparators considered in this section, and motivated the formalism presented in this paper.

\paragraph{Comparing an optical oscillator to an RF oscillator.} The RF oscillator being considered an accurate oscillator, the comparator output reads:
\begin{equation}
	\Cbi{\RFt}{\OOt} = \rr_{\OOt,\RFt}.
\end{equation}
This comparator can be implemented with an optical frequency comb, which is able to measure optical-to-RF frequency ratios, and from there one can compute the required RFR.

\paragraph{Comparing two optical oscillators with a frequency comb.} Given that the two optical oscillators are generally not accurate, this comparator must incorporate an auxiliary accurate oscillator, such as an RF reference. The output of the comparison can be generated from an optical-to-optical frequency ratio measurement and an optical-to-RF frequency ratio measurement, or equivalently from two optical-to-RF frequency ratio measurements, according to:
\begin{equation}
	\label{eq:COOOO}
	\Ctri{\OOt_1}{\OOt_2}{\RFt} = \rr_{\OOt_2,\OOt_1}(1+\rr_{\OOt_1,\RFt}) = \rr_{\OOt_2,\RFt} - \rr_{\OOt_1,\RFt}.
\end{equation}
All these quantities can be measured with an optical frequency comb. The second expression shows that this comparator can also be considered as two consecutive bipartite comparators, for which both $\rr_{\OOt_2,\RFt}$ and $\rr_{\OOt_1,\RFt}$ would be published.

It should be noted that, in line with comments in section \ref{sec:comparatormodels}, as neither $\OO_2$ nor $\OO_1$ are accurate, the comparator output must depend on an additional accurate auxiliary oscillator, even though the frequency comb could in principle measure the frequency ratio $\rr_{\OOt_2,\OOt_1}$ directly. Again, this trade-off is  necessary to accommodate both ratio measurements and beat frequency measurements within the same formalism.

\paragraph{Comparing an optical oscillator with an optical clock.} An optical clock can be considered as an optical oscillator, in which case the comparator output is given by equation~(\ref{eq:COOOO}). However, this comparator can also be written as a bipartite comparator considering that a clock is an accurate oscillator. In this configuration, the output of the comparator reads:
\begin{equation}
	\label{eq:COOOC}
	\Cbi{\clock}{\OOt} = \rr_{\OOt,\clock}.
\end{equation}
This output can be computed from the optical-to-optical frequency ratio $\rho_{\OOt,\clock}$, measured with a frequency comb. Internally, the comb's operator may still use an RF reference to produce $\rr_{\OOt,\clock}$, but this oscillator is not exposed to the network, and the comparator output~(\ref{eq:COOOC}) does not depend on the frequency of the RF reference, unlike when using equation~(\ref{eq:COOOO}).

\paragraph{Comparing a microwave clock to an RF oscillator.} Here, both oscillators are accurate, such that the comparator output can be indifferently set to $\Cbi{\RFt}{\clock} = \rr_{\clock, \RFt}$ or $\Cbi{\clock}{\RFt} = \rr_{\RFt,\clock}$, respectively. This quantity is the relative frequency difference between the RF oscillator and the clock. It is the quantity routinely produced by microwave fountains clocks, and usually referred to as ``the relative error of the RF oscillator''. With a proper choice of NFR (see section~\ref{sec:rrf}), its average is used to calibrate the TAI.

\paragraph{Systematic correction} The correction of the systematic effects of a clock can be modeled by a bipartite comparator in which one of the oscillators is the physical uncorrected output (u.o.) of the clock, and the other one the \emph{a posteriori} corrected output (c.o.). The corrected output can be considered accurate, such that choosing the NFR $\rho^0_{\textrm{\tiny c.o., u.o.}} = 1$, the comparator output $\Cbi{\textrm{\tiny c.o.}}{\textrm{\tiny u.o.}}$$=\rr_{\textrm{\tiny u.o.},{\textrm{\tiny c.o.}}}$, is the opposite of the relative frequency correction.

\section{Generic operation of the clock network}
\label{sec:computingratios}

\subsection{Computing frequency ratios}

We now aim at expressing the frequency ratio between two arbitrary remote oscillators in the network, one of which must be accurate~\footnote{In this formalism, comparing to an inaccurate oscillator $0$ requires either that an accurate oscillator is connected to $0$ by a bipartite comparator, or that only the stability of the frequency ratio is relevant. An example is proposed at the end of appendix~\ref{sec:usecase}.}, as a function of the individual outputs of the comparators connecting these oscillators. For this purpose, we have to rewrite equation~(\ref{eq:rhon0}) in terms of RFRs.

The frequency ratio $\rho_{n,0}$ between an arbitrary oscillator $n$ and an accurate oscillator $0$ connected by a series of comparators is constructed by induction, using the relation:
\begin{equation}
	\label{eq:rhoi0recur}
	\rho_{i,0} = \rho^0_{i,i-1}\rho_{i-1,0} + \Ctri{i-1}{i}{x_i}\,\rho^0_{i,x_i}\rho^0_{x_i,0}\left(1+\rr_{x_i,0}\right),
\end{equation}
where $x_i$ is the reference oscillator of the comparator linking the oscillators $i-1$ and $i$. This relation is derived in appendix~\ref{sec:proof} by applying the definition of the RFRs to the relation $\rho_{i,0} = \rho_{i,i-1}\rho_{i-1,0}$. It states that $\rho_{i,0}$ is constructed from $\rho_{i-1,0}$ by using the NFR $\rho^0_{i,i-1}$ between the linkage oscillators $i-1$ and $i$. Because $\rho^0_{i,i-1}$ differs from the actual frequency ratio $\rho_{i,i-1}$, a small correction is required, which is determined by the comparator output $\Ctri{i-1}{i}{x_i}$ and the connection between the accurate oscillators $\rr_{x_i,0}$. The recurrence relation~(\ref{eq:rhoi0recur}) can be explicitly solved, yielding:
\begin{eqnarray}
	\label{eq:rhon0lin}
	&& \rho_{n,0} = \left(\prod_{i=1}^n \rho^0_{i,i-1}\right)\left(1+\sum_{i=1}^n \R{i-1}{i}\right), \\
	&\textrm{with}\quad &
	\label{eq:R}
	\R{i-1}{i} = \Ctri{i-1}{i}{x_i}\frac{\rho^0_{i,x_i}\rho^0_{x_i,0}}{\prod_{k=1}^i\rho^0_{k,k-1}}\left(1+\rr_{x_i,0}\right).
\end{eqnarray}
These two equations state that the frequency ratio $\rho_{n,0}$ is approximated by the product of linkage NFRs, to which the terms $\R{i-1}{i}$ are small corrections (we have $\R{i-1}{i} \simeq \rr_{i,i-1}$, such that $|\R{i-1}{i}| \ll 1$) derived from the comparators outputs. More specifically, $\R{i-1}{i}$ is composed of three factors: the comparator output; a path-dependent constant, close to 1, that can be exactly calculated from known NFRs; and finally a correcting factor that depends on the RFR between the accurate oscillators $x_i$ and $0$. This latter RFR cannot be determined within this formalism because it is a non-local frequency ratio measurement. However, the RFRs between accurate oscillators are intentionally chosen small enough to be neglected. Hence:
\begin{equation}
	\label{eq:Rapprox}
	\R{i-1}{i} \simeq \Cshort{i-1}{i}{x_i}\frac{\rho^0_{i,x_i}\rho^0_{x_i,0}}{\prod_{k=1}^i\rho^0_{k,k-1}}.
\end{equation}
This approximation yields an error of the frequency ratio $\rho_{n,0}$ on the order of $\rr_{i-1,i}\rr_{x_i,0}$. If the NFRs are chosen, for example, such that $|\rr_{i-1,i}| < 10^{-6}$, and $|\rr_{x_i,0}| < 10^{-13}$, the error on $\R{i-1}{i}$ is lower than $10^{-19}$.

Because $|\R{i-1}{i}| \ll 1$, double precision floating point arithmetic is sufficient to calculate $\R{i-1}{i}$ with a resulting relative error on the final frequency ratio lower than $10^{-16}\R{i-1}{i}$. Consequently, double precision is also sufficient to store the comparator outputs and to define the NFRs between accurate oscillators.

Arbitrary precision is eventually required to compute the leading product of the linkage \NFRs in equation~(\ref{eq:rhon0lin}). However, this product can be calculated once for a given pair of oscillators $(n,0)$, as opposed to calculated at the data sampling rate. The linkage \NFRs must therefore be numerically representable exactly, for example as rational numbers given by a pair of integer numbers for the numerator and denominator, or at least with the same precision as the target uncertainty of the clock comparison.

\subsection{Data exchange protocol}

The operation of the network consists of the following steps:
\begin{enumerate}
	\item Each comparator chooses its local NFRs, and uses them to produce the time series for the comparator output $\Cshort{i}{j}{x_i}$, which is published together with the NFRs.
	\item The data from the various comparators are collected. From these data, together with a sensible choice for the NFRs between the accurate oscillators, the various frequency ratios between the oscillators of the network can be computed with equations~(\ref{eq:rhon0lin}) and~(\ref{eq:Rapprox}).
\end{enumerate}

\section{Oscillator-centric choice of NFRs}
\label{sec:oscillatorcentric}

Observing equations~(\ref{eq:rhon0lin}) to~(\ref{eq:Rapprox}), one is tempted to simplify the NFRs using the chain rule:
\begin{equation}
	\label{eq:chain}
	\rho^0_{a,c} = \rho^0_{a,b}\,\rho^0_{b,c}.
\end{equation}
However this relation is not  valid for all possible choices of \NFRs. For instance, applying the chain rule, both numerator and denominator in equation~(\ref{eq:Rapprox}) would reduce to $\rho^0_{i,0}$, but following different paths: through the connection of accurate oscillators for the numerator, or through the chain of compared oscillators for the denominator. Because the \NFRs along these two paths are generally chosen independently, their contractions to $\rho^0_{i,0}$ are \emph{a priori} different. In other words, with an individual choice of \NFRs, the set of \NFRs is generally not transitive.

However, in the examples of comparators that we have seen in section~\ref{sec:comparatorexamples}, all the comparators -- save for the frequency beatnote -- are actually able to produce their output for any choice of their local \NFRs. Consequently, the network operators may collectively decide to withdraw the freedom of choice for the comparator local \NFRs, and agree on a transitive set of \NFRs, for which the chain rule is applicable. In this section, we describe the consequences of such a choice, and propose an explicit construction, whose interpretation is centered on oscillators..

\subsection{Computing frequency ratios with oscillator-centric NFRs}

Assuming the chain rule~(\ref{eq:chain}) applies to all \NFRs, equations~(\ref{eq:rhon0lin}) to~(\ref{eq:Rapprox}) can be simplified to:
\begin{eqnarray}
	\label{eq:rhoRrr}
	&& \rr_{n,0} = \sum_{i = 1}^n \R{i-1}{i},\\
	\label{eq:Ri}
	& \textrm{with} \quad & \R{i-1}{i} = \Ctri{i-1}{i}{x_i}\left(1+\rr_{x_i,0}\right) \simeq \Cshort{i-1}{i}{x_i}.
\end{eqnarray}
With a transitive choice of NFRs, the reduced frequency ratios between remote oscillators  are therefore computed by summing the individual comparator outputs $\Cshort{i-1}{i}{x_i}$, thus allowing for a more simple data analysis.

Given that the comparator output $\Cshort{i-1}{i}{x_i}$ is either a RFR~(\ref{eq:Cij2}), or a difference of RFRs~(\ref{eq:Cij3lin}), equations~(\ref{eq:rhoRrr}) and~(\ref{eq:Ri}) simply state that remote RFRs are obtained by adding or subtracting local RFRs. This is similar to the customary procedure to compare clocks by satellite evoked in section~\ref{sec:network}, except that in the framework of this paper, the signs of the RFRs in the combination must be carefully chosen in order to ensure the accuracy of the comparison. This specificity stems from the use of frequency offsets and inaccurate oscillators in the network.

Note that the direct proof of relations~(\ref{eq:rhoRrr}) and~(\ref{eq:Ri}) is straightforward, combining $\rr_{n,0} = \sum_{i = 1}^n  \rr_{i,0} - \rr_{i-1,0}$ and $ \rr_{i,0} - \rr_{i-1,0} = (\rr_{i,x_i} - \rr_{i-1,x_i})(1+\rr_{x_i,0})$.

\subsection{Data exchange protocol with oscillator-centric NFRs}

The operation of the network consists in the following steps:
\begin{enumerate}
	\item Prior to operation, the network collaboration agrees on a network-wide consistent set of NFRs that is chosen both transitive and such that the RFR between accurate oscillators are small enough.
	\item During operation, each comparator produces the time series for the comparator output $\Cshort{i}{j}{x_i}$, which is published.
	\item The data from the various comparators are collected. The various reduced frequency ratios between the oscillators of the network can be computed by summing the shared data along a given path with:
	\begin{equation}
		\label{eq:RC}
		\rr_{n,0} \simeq \sum_{i = 1}^n \Cshort{i-1}{i}{x_i}.
	\end{equation}
\end{enumerate}

\subsection{Nominal frequencies}
\label{sec:rrf}

We now propose a possible method to effectively produce a transitive set of NFRs, \emph{i.e.} satisfying~(\ref{eq:chain}) for all NFRs. It relies on the fact that the transitivity assumption~(\ref{eq:chain}) is equivalent to assuming that for each oscillator $i$, there exists a numerical constant $\nuz_i$, hereafter called the ``nominal frequency'' of oscillator $i$, such that:
\begin{equation}
	\label{eq:defnu0}
	\rho^0_{i,j} = \frac{\nuz_i}{\nuz_j}
\end{equation}
for all pairs of oscillators $(i, j)$. Here, we consider the quantities $\nuz_i$ as dimension-less, exactly known numerical constants that are only ever to be used with the intent to compute RFRs. It is therefore not necessary to attribute a unit to them, from which it follows that the quantity $\nu_i - \nuz_i$, never used in this paper, is left undefined.

The point of view adopted in this choice of NFRs is oscillator-centric, in the sense that the NFRs are determined by attributing a label $\nuz_i$ to each oscillator in the network. This contrasts with the more general, comparator-centric point of view introduced in section~\ref{sec:networkcomponents} that consists in attributing NFRs to each comparator in the network.

For a given set of transitive \NFRs, the choice of nominal frequencies is not unique. Indeed, they can be collectively multiplied by the same number without altering the \NFRs. Making use of this possibility, one can ensure that the nominal frequency $\nuz_i$ is close to the numerical value of the frequency $\nu_i$ in a given frequency unit, such as the SI Hz. Conversely, one can use an approximate value of the oscillators' frequencies expressed in a specific frequency unit in order to practically define a transitive set of \NFRs.
For instance, for an atomic clock whose clock transition is promoted to secondary representation of the SI second, the nominal frequency can be chosen as the numerical value of the transition frequency recommended by the Comit\'e International des Poids et Mesures (CIPM). The nominal frequency of an RF source can be set to its regular nominal frequency. For infrared optical oscillators transmitted through optical fiber links, the nominal frequency can be chosen for example as the laser frequency rounded at the 100~MHz level, or as the central frequency of the telecom channel if close enough.

This choice of nominal frequencies duly satisfies $|\rr_{i,j}| < 10^{-6}$ for linkage oscillators and $|\rr_{i,j}| < 10^{-13}$ for pairs of accurate oscillators, as well as the requirement that $\rho^0_{i,j} = 1$ for frequency offsets and beatnotes, with $\nuz_j = \nuz_i$ for the corresponding lasers.

Choosing the nominal frequencies of clocks such that they are independent of the network architecture (such as the CIPM recommended values) also means that the RFRs $\rr_{n,0}$ between remote clocks acquire an intrinsic meaning: they quantify the relative discrepancy between the measured ratio and the recommended ratio, making them easier to mentally handle and compare than  frequency ratios with 19 significant digits. The reduced frequency ratios can therefore be considered as the final output of the network, and they can be directly produced by equation~(\ref{eq:RC}), exclusively using double precision arithmetic. This procedure is already in place in the BIPM's circular T: clocks used to calibrate TAI report their measured value of $\rr_{\RFt,\clock}$ (averaged over a given period), where RF is the reference generating the local UTC(k) used as a pivot to TAI, and the NFR $\rho^0_{\RFt,\clock}$ is defined according to the prescription given in this section.

\section{The transfer beat notation}
\label{sec:TBF}

In this section, we aim at introducing an equivalent notation for the comparator output called the ``transfer beat''. Its purpose is to more intuitively connect the comparator output to the numerical output of the measurement devices, by making use of the concept of nominal frequency introduced in section~\ref{sec:rrf}. The transfer beat notation underlies a set of \NFRs that is neither purely comparator-centric nor purely oscillator-centric. This notation requires that nominal frequencies for at least the \emph{accurate} oscillators are defined, i.e. the NFRs between accurate oscillators are transitive. However, the local NFRs of the comparators can be independently chosen.

The transfer beat notation interprets the formalism developed in this paper as a generalization of the transfer oscillator technique employed with optical frequency combs~\cite{ste02a, tel02b}. The analogy between these concepts is developed in detail in appendix~\ref{sec:transferoscillator}.

\subsection{Measured frequencies}
\label{sec:measuredfrequencies}

The formalism developed in this paper is exclusively expressed in terms of dimension-less ratios between the physical frequencies $\nu_i$ of the oscillators composing the network. Therefore, it is independent of the definition of a frequency unit. We may nevertheless use the concept of nominal frequency to define the numerical outcome $\hat f$ of the measurement of a signal with frequency $f$, referenced to an accurate oscillator $x$, by:
\begin{equation}
	\hat f \equiv \frac{f}{u_x} \qquad \textrm{with} \quad \u_x = \frac{\nu_x}{\nuz_x},
\end{equation}
where $\nu_x$ is the actual physical frequency of the accurate oscillator, and $\nuz_x$ its chosen dimensionless nominal frequency. $\hat f$ is thus an alternative notation for the frequency ratio $f/\nu_x$. It is useful when, for instance, $x$ is an RF local oscillator referencing a frequency counter. In this case, $\hat f$ conveniently matches the number output by the counter, provided $\nuz_\RFt$ coincides with the expected reference frequency of the counter  (\emph{e.g.} $\nuz_\RFt = 10^7$ if the counter expects to be referenced to a 10~MHz reference). It also follows that $\hat f$ is independent of the local scaling applied to the RF reference, \emph{i.e.} of a multiplicative factor applied to both $\nu_\RFt$ and $\nuz_\RFt$.

This notation may also be used to represent the frequency ratio between an optical oscillator $j$ and an optical clock $i$. However, one should keep in mind that although $\hat \nu_j \equiv \nu_j/u_i$ may look like a quantity expressed in Hz, it is in reality nothing but a representation of the pure optical-to-optical frequency ratio $\rho_{j,i}$. In this view, it may be useful to interpret the frequency $\u_x$ as a frequency unit, for which the numerical value of the frequency $\nu_x$, when expressed in this unit, is $\nuz_x$ exactly. The numerical value of $f$ in $\u_x$ is $\hat f$.

Finally, the notation $\hat f$ is ambiguous because it does not explicitly mention the oscillator used as a reference. However, in this paper, this oscillator will always be the reference oscillator of the comparator in which $\hat f$ is measured.

\subsection{Computing frequency ratios using transfer beats}
\label{sec:TBFdef}

The transfer beat associated with the comparator linking the oscillator $i$ to the oscillator $j$ is defined as:
\begin{equation}
	\label{eq:rrtotbf}
	\ff{i}{j} \equiv -\rr_{i,j}\nu_j = \nu_j - \rho^0_{j,i}\nu_i,
\end{equation}
generalizing the concept of beatnote $\Delta f = \nu_j - \nu_i$. The transfer beat satisfies:
\begin{equation}
	|\ff{i}{j}| \ll |\nu_j|,
\end{equation}
and the transfer beats in the forward and backward directions are linked by:
\begin{equation}
	\ff{i}{j} = -\rho^0_{j,i} \fb{i}{j}.
\end{equation}
The comparator output $\Ctri{i}{j}{x}$ is related to the transfer beat through:
\begin{equation}
	\Ctri{i}{j}{x} = \frac{\ff{i}{j}}{\nu_x}\rho^0_{x,j} = \frac{\hff{i}{j}}{\nuz_x}\rho^0_{x,j},
\end{equation}
where $\hff{i}{j}$ is the transfer beat measured against the comparator's reference oscillator $x$, whose agreed upon nominal frequency is $\nuz_x$. The expression~(\ref{eq:R}) of $\R{i-1}{i}$ can then be rewritten in terms of the measured transfer beats:
\begin{equation}
	\label{eq:tbfRunapprox}
	\R{i-1}{i} = \frac{\hff{i-1}{i}}{\nuz_0}\frac{1}{\prod_{k=1}^i\rho^0_{k,k-1}}\left(1+\rr_{x_i,0}\right),
\end{equation}
which, like previous cases, can be approximated by neglecting $\rr_{x_i,0}$:
\begin{equation}
	\label{eq:tbfRapprox}
	\R{i-1}{i} \simeq \frac{\hff{i-1}{i}}{\nuz_0}\frac{1}{\prod_{k=1}^i\rho^0_{k,k-1}},
\end{equation}
with the approximation error being proportional to the value of the transfer beat. Using the transfer beat notation, neither the measured transfer beat nor the approximation~(\ref{eq:tbfRapprox}) of $\R{i-1}{i}$ depend on the choice of the local NFRs $\rho^0_{i,x}$ and $\rho^0_{i-1,x}$. It is therefore unnecessary to define or share these quantities.

\subsection{Data exchange protocol with transfer beats}

The operation of the network consists of the following steps:
\begin{enumerate}
	\item Prior to operation, the network collaboration agrees on a network-wide consistent set of nominal frequencies for the accurate oscillators~\footnote{More precisely, for the accurate oscillators that act as references for the comparators. The nominal frequency of the other accurate oscillators can be defined later when calculating the frequency ratios.}. If a specific frequency unit, for instance the SI Hz, is agreed upon in order to define these nominal frequencies, as proposed in section~\ref{sec:rrf}, it is not necessary to publish their actual values, because the required numerical precision is moderate compared to the uncertainty of the absolute frequency of clocks found in the literature (see section~\ref{sec:computingratios}).
	\item During operation, each comparator chooses a linkage NFR $\rho^0_{j,i}$, and produces the time series for the measured transfer beats $\hff{i}{j}$, which is published together with $\rho^0_{j,i}$.
	\item The data from the various comparators are collected. The various reduced frequency ratios between the oscillators of the network can be computed with equations~(\ref{eq:rhon0lin}) and~(\ref{eq:tbfRapprox}).
\end{enumerate}

\subsection{Generating the transfer beats}

The measured transfer beats are, within a known multiplicative constant, identical to the comparator outputs. We nevertheless explicitly write them here to outline their physical significance.

\paragraph{For a bipartite comparator,} whose reference oscillator is $i$, the measured transfer beat reads
\begin{equation}
	\label{eq:tfbbipartite}
	\hff{i}{j} = \left(\rho_{j,i} - \rho^0_{j,i}\right)\nuz_i = \hat\nu_j - \rho^0_{j,i}\nuz_i  \quad \textrm{if $i$ is accurate,}
\end{equation}
where $\hat \nu_j = \nu_j/\u_i$ is the frequency of $j$ measured against the accurate oscillator $i$, \emph{i.e.} the numerical value of the frequency $\nu_j$ in the unit $\u_i$. The choice of NFR $\rho^0_{j,i}$ just enters as a constant offset that effectively reduces the number of significant digits required to accurately represent the transfer beat.

As a practical example, we consider the comparison between an optical oscillator $\OO$ and an RF oscillator with an optical frequency comb. The transfer beat $\hff{\RFt}{\OOt}$ can be computed from the comb measurement output for an arbitrary NFR $\rho^0_{\OOt,\RFt}$ by using equation~(\ref{eq:tfbbipartite}) in conjunction with:
\begin{equation}
	\label{eq:hatnucomb}
	\hat\nu_\OOt = m_{\OOt} \hfrep + \hat \delta_{\OOt},
\end{equation}
where $m_{\OOt}$ is the index of the comb tooth beating with the optical oscillator $\OO$ producing the beatnote $\delta_{\OOt}$ (including the carrier envelope offset), and $\frep$ is the comb's repetition rate. Instead of publishing the NFR $\rho^0_{\OOt,\RFt}$, the comb operator may directly publish the product $\rho^0_{\OOt,\RFt}\nuz_\RFt$, which is independent of the local scaling applied to the RF oscillator before it is fed to the measurement apparatus.

\paragraph{For a tripartite comparator,} the measured transfer beat derives directly from the definition~(\ref{eq:rrtotbf}):
\begin{equation}
	\hff{i}{j} = -\rr_{i,j}\hat \nu_j = \hat\nu_j - \rho^0_{j,i}\hat \nu_i,
\end{equation}
where $\hat \nu_i$ and $\hat\nu_j$ are the frequencies of $i$ and $j$ measured against the accurate reference $x$.

As a practical example, we consider the comparison between two optical oscillators $\OO_1$ and $\OO_2$ with a frequency comb, using an RF reference. The transfer beat can be computed from the comb's output for an arbitrary NFR with:
\begin{equation}
	\label{eq:gentbcomb}
	\hff{\OOt_1}{\OOt_2} = \hat\nu_{\OOt_2} - \rho^0_{\OOt_2,\OOt_1}\hat\nu_{\OOt_1},
\end{equation}
where the quantity $\hat \nu_\OOt$ can be computed from the comb's output with equation~(\ref{eq:hatnucomb}).

\section{Formalism error}
\label{sec:comparing}

The formalism described in this paper introduces an error on the computation of frequency ratios. Each comparator is responsible for a relative error of the determined frequency ratio $\rho_{n,0}$ on the order of $\rr_{j,i}\rr_{x,0}$ where $i$, $j$, and $x$ are the oscillators involved in the comparator and $0$ is the first oscillator in the remote frequency ratio. In the case of genuinely tripartite comparators, for example a beat frequency measurement, this error is dictated by physical parameters and is unavoidable irrespective of formalism. In the case of bipartite comparators however, we deliberately choose to approximate the directly measurable ratio $\rho_{j,i}$, thereby introducing an error -- this is the price for being able to treat all comparators in the same way. In this section, we present a detailed discussion of this error.

In the examples we have discussed earlier, the linkage RFRs $\rr_{i,j}$ can generally be lower than $10^{-6}$. RFRs that involve optical oscillators are indeed on the order of $\eo \simeq 10^{-6}$, and RFRs handled with the transfer oscillator method (see appendix~\ref{sec:transferoscillator}) are as large as $\eb \simeq 10^{-6}$. The NFRs between accurate oscillators can be typically chosen such that the corresponding RFRs are lower than $10^{-13}$, with RFRs between RF references on the order of $\er \simeq 10^{-13}$, and down to below $\ec \simeq 10^{-15}$ for RFRs between clocks. With these bounds, the comparison of oscillators in the network is altered by a maximum relative error of $N\times 10^{-19}$ on the computation of remote frequency ratios, where $N$ is the number of comparators in the chain (see appendix~\ref{sec:overallerror} for more details). We stress at this point that the numerical values chosen for the bounds are only an example, roughly aligned with current technical capability in the European clock comparison network.

\begin{table*}
\renewcommand{\arraystretch}{1.2}
\begin{center}
\scalebox{0.8}{
\hspace{-1cm}
\begin{tabular}{lclc}
 \hline
 \hline
 Comparison & Type & Interpretation of the transfer beat  $\hff{i}{j}$ or of the comparator output $\Delta_{i\rightarrow j}$& Error  \\
 \hline
 Optical oscillator vs. RF oscillator & Bipartite & $\hff{\RFt}{\OOt}$ is the frequency of the optical oscillator in $\u_\RFt$, reduced by $\rho_{\OOt,\RFt}\nuz_\RFt$ & $\eo\er$\\
 Microwave clock vs. RF oscillator & Bipartite & $\Cbi{\RFt}{\clock}$ is the RFR $\rr_{\clock,\RFt}$ \emph{i.e.} the RF (\emph{e.g} H-Maser) relative error & $\ec\er$\\
 Optical oscillator vs. optical oscillator & Tripartite* & Sum of two bipartite optical to RF comparisons & $\eo\er$ \\
 Optical oscillator vs. optical oscillator & Tripartite & Transfer oscillator technique with combs~\cite{ste02a,tel02b}, appendix~\ref{sec:transferoscillator} & $\eb\er$ \\
 Optical oscillator vs. optical clock  & Bipartite & $\hff{\clock}{\OOt}$ is the frequency of the optical oscillator in $\u_\clock$, reduced by $\rho_{\OOt,\clock}\nuz_\clock$ & $\eo\ec$\\
 Beatnote between optical oscillators  & Tripartite & The transfer beat is the counter output numerical value & $\eo\er$ \\
 \hline\hline
\end{tabular}}
\end{center}
\caption{\label{tab:Ri}Example of comparators in a network of optical and microwave clocks connected by optical fiber links. The second column tells if the comparator is bipartite or tripartite. The third column proposes a physical interpretation of the transfer beat generated by the comparator, or of the comparator output $\C{i}{j}$. The comparison between two optical oscillators is shown first in its general form, and then in two special cases where the transfer oscillator method is used (as detailed in appendix~\ref{sec:transferoscillator}), and where one of the oscillators is accurate. The last column gives the order of magnitude of the error introduced by the formalism, for which the quantities $\epsilon$ quantify the contribution of a given type of oscillator to the RFRs, as defined in the main text. The tripartite comparator indicated with a star is reducible to two bipartite comparators in series. This is not the case for the other two tripartite comparators for which the frequency ratio between the oscillators and the accurate reference is not available due to the lack of frequency comb for beatnote measurements, or the choice not to exploit the measurement of the repetition rate in the case of the transfer oscillator technique.}
\end{table*}

Table~\ref{tab:Ri} summarizes the error introduced by the different types of comparison methods we have described so far. It is noteworthy that most of the approximation errors are on the order of $\eo$, reduced by an extra factor. This error is negligible compared to the current and foreseeable uncertainty of optical clocks, but it can nevertheless easily be reduced by orders of magnitude by controlling the optical oscillator frequencies so that they remain as close as possible to their nominal values (or vice versa), and decreasing the beat frequencies between the optical oscillators accordingly. This process is only limited by the long term frequency fluctuations of the optical oscillators~\cite{Hagemann:14}.

This summary shows that a proper choice of NFRs is essential to reach a low  error in the final computation of frequency ratios. We show in appendix~\ref{sec:changeNFR} that for tripartite comparators, there is generally no direct link between the comparator output for two different choices of \NFRs. Hence, these choices are not equivalent.
The considerations driving the choice of \NFRs are threefold:
\begin{itemize}
	\item \emph{Approximation error:} Because the error introduced by the formalism depends on the mismatch between the oscillators' frequency ratio and the \NFRs, one should ensure that the \NFRs are close enough to the actual frequency ratios for this error to be negligible.
	\item \emph{Enabling measurements:} Some choice of \NFRs can circumvent possible errors in the comparator, or make the measurement possible at all. In this paper, we have seen that prescribing  \NFRs for frequency beatnotes and offsets is required to express the comparator output in terms of measured quantities. Also, the transfer oscillator scheme~\cite{ste02a, tel02b} historically emerged from this concern. In this respect, allowing each comparator to individually choose an appropriate value for its local \NFRs is required (see appendix~\ref{sec:transferoscillator}).
	\item \emph{Comparator type:} When possible, exposing a comparator as a bipartite comparator relaxes the choice of \NFRs since this choice can be amended in the data post-processing. We also show in appendix~\ref{sec:exactfr} that selecting bipartite comparators preserves enough information about the comparison to enable the calculation of exact frequency ratios, \emph{i.e.} the formalism error can be eliminated.
\end{itemize}

If several choices of \NFRs are suitable, selecting a comparator-centric set of \NFRs yields flexibility in the network: each comparator can select its own \NFR, independently of the other comparators of the network, also avoiding the necessity of an \emph{a priori} agreement on nominal frequencies. On the other hand, selecting an oscillator-centric set of nominal frequencies eases the data analysis, for the reduced frequency ratios can be directly calculated by simply adding the shared comparator outputs with double precision arithmetic. Furthermore, with the oscillator centric choice, the comparator outputs do not expose internal parameters of the comparators, such as the mode numbers $m$ of combs.

\section{Conclusion}

In this paper, we have shown that the experimental outcomes of a multi-partite comparison of atomic clocks can be split into local, path-independent quantities that can be produced as the output of each comparator connecting the oscillators of the network. The comparator outputs can be cast into a form based on locally determined frequency ratios,
or equivalently into the form of a transfer beat. Sharing these quantities is sufficient to calculate the frequency ratio between any pair of oscillators within the network using a generic, scalable procedure that can be implemented using double precision arithmetic. The key ingredient of the formalism is the availability of \emph{a priori} estimates of the frequency ratios between locally compared oscillators, as well as between a set of oscillators designated as accurate references accessible to all comparators. The formalism yields only approximate values of remote frequency ratios, but the error can be made significantly smaller than the current and foreseeable uncertainty of the best optical clocks.

\begin{acknowledgments}
This work was financially supported by the UK Government Department for Business, Energy and Industrial Strategy as part of the National Measurement System Programme; the European Metrology Programme for Innovation and Research (EMPIR) projects 15SIB03 OC18, 15SIB05 OFTEN and 18SIB05 ROCIT. These projects have received funding from the EMPIR programme co-financed by the Participating States and from the European Union’s Horizon 2020 research and innovation programme.

JK acknowledges helpful discussions with members of the Time and Frequency group at NPL and thanks Charles Baynham and Rich Hendricks for critical review of the manuscript prior to submission. EB and SK acknowledge valuable comments from Alexander Kuhl and Thomas Waterholter.
\end{acknowledgments}

\appendix

\section{Proof of equation~(\ref{eq:rhoi0recur})}
\label{sec:proof}

In this appendix, we derive the induction relation~(\ref{eq:rhoi0recur}):
\begin{eqnarray}
	&   & \rho_{i,0} - \rho^0_{i,i-1}\rho_{i-1,0} \nonumber\\
	& = & \rho_{i,x_i}\rho_{x_i,0} - \rho^0_{i,i-1}\rho_{i-1,x_i}\rho_{x_i,0} \label{eq:proof1}\\
	& = & \left(\rho_{i,x_i}\rho^0_{x_i,i} - \rho_{i-1,x_i}\rho^0_{x_i,i-1}\right)\rho^0_{i,x_i}\rho_{x_i,0} \label{eq:proof2}\\
	& = & \Ctri{i-1}{i}{x_i}\,\rho^0_{i,x_i}\rho_{x_i,0} \label{eq:proof3}\\
	& = & \Ctri{i-1}{i}{x_i}\,\rho^0_{i,x_i}\rho^0_{x_i,0}\left(1+\rr_{x_i,0}\right) \label{eq:proof4}
\end{eqnarray}
where~(\ref{eq:proof1}) uses the transitivity of frequency ratios; (\ref{eq:proof2}) uses the transitivity of the local NFRs, assumed in section~\ref{sec:comparatormodels}; (\ref{eq:proof3}) uses the definition~(\ref{eq:Cij3lin}) of the comparator output; and (\ref{eq:proof4}) uses the definition~(\ref{eq:rr}) of the RFRs.

\section{Connection with the transfer oscillator method}
\label{sec:transferoscillator}

The definition of the transfer beats introduced in section~\ref{sec:TBFdef} is inspired by the transfer oscillator technique employed with optical frequency combs~\cite{ste02a, tel02b}. It was introduced to mitigate the need for high-bandwidth phase-locking of the comb parameters to a frequency reference in frequency ratio measurements between an optical clock and a microwave clock, in frequency ratio measurements between optical clocks, or during the frequency stability transfer between optical oscillators.
This is accomplished by RF- or post-processing of the carrier-envelope-offset (CEO) beat signal and the beat signals between the comb and subject continuous laser field resulting in a so-called transfer-beat that is free of the CEO frequency $f_\textrm{\tiny CEO}$ and repetition rate, and thus of their fluctuations. Since only RF beat frequencies which are offsets to optical frequencies are compared to an RF reference by means of a frequency counter, the error due to the RF reference is suppressed by a factor given by the ratio between the optical and RF beat frequency, resulting in an error similar to the approximation error in equation~(\ref{eq:tbfRapprox}).

Here, we show the analogy between the computation of the frequency ratio between two optical clocks $0$ and $1$ using the transfer oscillator concept, and the generic network formalism we developed in this paper. The transfer oscillator technique is based on writing the frequency of each oscillator $i$ as a small offset $\delta_i$ (in practice $f_\textrm{\tiny CEO} + f_{i-\mathrm{comb}})$ to a nominal frequency written as an integer multiple $m_i$ (the comb mode number) of the comb's repetition rate $\frep$:
\begin{equation}
    \label{eq:rubberband}
	\nu_i = m_i\frep + \delta_i.
\end{equation}
One then forms the transfer beat
\begin{equation}
	\label{eq:fftransosc}
	\ff{0}{1} = \nu_1 - \frac{m_1}{m_0}\nu_0 = \delta_1 - \frac{m_1}{m_0}\delta_0,
\end{equation}
which uses the ratios of the mode numbers $m_1/m_0$ as a linkage NFR $\rho^0_{1,0}$. The transfer beat does not involve $\frep$, thus avoiding any noise its measurement may have added. The independence of the transfer beat from the repetition rate is analogous to the irrelevance of the unit in which the nominal frequencies of equation~(\ref{eq:chain}) are expressed.

Computing the frequency ratio $\rho_{1,0}$ then involves a measurement $\hff{0}{1}$ of the transfer beat against a local oscillator $x$, typically an RF reference, together with a good enough \emph{a priori} estimate $\rho^0_{0,x}$ of the frequency ratio between the clock 0 and the local oscillator $x$, or, said otherwise, a good enough estimate $\nuz_0 \equiv \rho^0_{0,x}\nuz_x$ of the frequency $\hat \nu_0$ in units of $\u_x$:
\begin{equation}
	\label{eq:transosc}
	\rho_{1,0} = \rho^0_{1,0}  + \frac{\ff{0}{1}}{\nu_0} = \rho^0_{1,0} + \frac{\hff{0}{1}}{\hat \nu_0} \simeq \rho^0_{1,0} + \frac{\hff{0}{1}}{\nuz_0}.
\end{equation}
The approximation error in equation~(\ref{eq:transosc}) is $\frac{\hff{0}{1}}{\hat \nu_0}\rr_{x,0}$.
We note $\eb$ the order of magnitude of $|\ff{0}{1}/\nu_0|$. For $\ff{0}{1}$ in the 100~MHz range, $\eb \simeq 10^{-6}$, which yields a relative error on the order of $10^{-19}$ on $\rho_{1,0}$,  assuming $|\rr_{x,0}| \simeq 10^{-13}$. It is the analog of the approximation error~(section~\ref{sec:comparing}) in the network formalism.

The availability of the \emph{a priori} estimate $\rho^0_{0,x}$, independently measured with appropriate equipment and transmitted to the comb's operator \emph{e.g.} via the realization of a frequency unit, is a critical ingredient to compute the frequency ratio $\rho_{1,0}$ without relying on the information carried by $\frep$. Without this estimate, the comb's operator could still measure the frequency $\hat \nu_0$ that appears in~(\ref{eq:transosc}) with the comb itself, using~(\ref{eq:hatnucomb}), and thus fall back on the expression of $\rho_{1,0}$ that directly follows from~(\ref{eq:rubberband}):
\begin{equation}
	\label{eq:absfratio}
	\rho_{1,0} = \frac{\hat \nu_1}{\hat \nu_0} =  \frac{m_1\hfrep + \hat \delta_1}{m_0\hfrep + \hat \delta_0}.
\end{equation}
However, in this RF reference-free optical-to-optical frequency ratio measurement (analogous to considering the comb as a reference-free bipartite comparator in our network formalism), the relative added noise on the measurement of $\frep$ that does not adhere to the elastic tape model of equation~(\ref{eq:rubberband}) is only rejected by a factor $\delta_i/\nu_i \sim 10^{-6}$.

The required \emph{a priori} estimate $\rho^0_{0,x}$ for the transfer oscillator method is analogous to the required \emph{a priori} NFRs between accurate oscillators in our network formalism. In the latter, this requirement does not originate from lack of an appropriate apparatus, but from the non-locality of the oscillators $x_i$ and $0$, which forbids a direct measurement of $\rho_{x_i,0}$.

The transfer oscillator technique can be implemented with the formalism presented in this paper, by noticing that equation~(\ref{eq:fftransosc}) is compatible with equation~(\ref{eq:gentbcomb}) when choosing $\rho^0_{\OOt_2,\OOt_1} = m_{\OOt_2}/m_{\OOt_1}$. With this choice of NFR, any added noise $\delta\frep$ on the measurement of $\frep$ is canceled. Instead, choosing the NFR as close as possible to the actual frequency ratio $\rho_{\OOt_2,\OOt_1}$ minimizes the formalism  while the contribution from unsuppressed uncorrelated repetition rate noise increases.

The break-even point between these two choices is reached when $\left|\delta\frep/\frep\right| \simeq \left|\rr_{x,0}\right|$. This illustrates a final analogy between the transfer oscillator scheme and our network formalism. A frequency comb for which $\left|\delta\frep/\frep\right| > \left|\rr_{x,0}\right|$ would not be suitable to precisely measure the local optical-to-RF frequency ratios $\rr_{\OOt_1, x}$ and $\rr_{\OOt_2, x}$. However, even though these two quantities are not individually accessible without added noise, their difference, that is to say the comparator output, rejects this noise. This is analogous to frequency beatnotes, for which the comparator output is experimentally accessible without any device suitable to measure individual frequency ratios. The transfer oscillator technique is thus another example showing that implementing a tripartite comparator together with a proper choice of linkage NFR can be useful in order to circumvent technical limitations of a comparator.

\section{Change of \NFRs}
\label{sec:changeNFR}

In this section we investigate if there exists a transformation between the comparator outputs produced for different choices of local \NFRs.

We note $\C{i}{j}$ the comparator output generated with a choice $\rho^0_{i,j}$ for the linkage NFR, and $\Cp{i}{j}$ the output for another choice $\rho'^0_{i,j}$, and try to establish a relation between $\C{i}{j}$ and $\Cp{i}{j}$.

\paragraph{For a bipartite comparator,} we have the simple relation
\begin{equation}
	\Cbip{i}{j} = (\Cbi{i}{j}+1)\frac{\rho^0_{j,i}}{\rho'^0_{j,i}} - 1   \qquad \textrm{if $i$ is accurate},
\end{equation}
which can also be written with the transfer beat notation:
\begin{equation}
	\hfpf{i}{j} = \hff{i}{j} + (\rho^0_{j,i} - \rho'^0_{j,i})\,\nuz_i   \qquad \textrm{if $i$ is accurate}.
\end{equation}
As a consequence of these relations, it is possible to \emph{a posteriori} change the \NFR in order to reduce the magnitude of the comparator output, and thus reducing the error introduced by the formalism, if required. Therefore, the initial choice of \NFR does not entail any practical limitation.

\paragraph{For a tripartite comparator,} generating the comparator output requires the definition of two NFRs: the linkage NFR $\rho^0_{j,i}$, but also $\rho^0_{j,x}$ (or alternatively $\rho^0_{i,x}$, these three NFRs being linked by the local transitivity relation). From equation~(\ref{eq:CijT}), it is clear that $\rho^0_{j,x}$ is simply a multiplicative constant that scales the comparator output. Furthermore, the final frequency ratio $\rho_{n,0}$ is independent of this scaling because it is canceled by the factor $\rho^0_{i,x_i}$ in the expression~(\ref{eq:R}) of $R_{i-1,i}$. Therefore, $\rho^0_{j,x}$ can be indifferently set to any convenient value, that does not even have to be close to the actual frequency ratio $\rho_{j,x}$. We exploited this freedom in section~\ref{sec:TBFdef}, in which the transfer beat reduces to the comparator output when setting $\rho^0_{x,j} = \nuz_x$.

On the contrary, the choice of the linkage NFR $\rho^0_{j,i}$ cannot be \emph{a posteriori} changed, because there is no exact numerical relation linking $\Ctrip{i}{j}{x}$ to $\Ctri{i}{j}{x}$ when this NFR changes. Indeed, they are related by:
\begin{equation}
	\Ctrip{i}{j}{x} = \Ctri{i}{j}{x}\frac{\rho^0_{j,i}}{\rho'^0_{j,i}} + \left(\frac{\rho^0_{j,i}}{\rho'^0_{j,i}} - 1\right)(1+\rr_{i,x}),
\end{equation}
which can also be written with the transfer beat notation:
\begin{equation}
	\label{eq:tfbprime}
	\hfpf{i}{j} = \hff{i}{j} + (\rho^0_{j,i} - \rho'^0_{j,i})\,\hat \nu_i.
\end{equation}
Changing the NFRs is therefore only feasible if an independent measurement of the frequency ratio between the oscillator $i$ against the accurate reference $x$ is available. However, if $i$ itself is also an accurate oscillator, neglecting $\rr_{i,x}$ (\emph{i.e.} replacing $\hat \nu_i$ with $\nuz_i$ in the transfer beat notation) yields a good enough approximate transformation between the comparator outputs.

For instance, if a transfer beat $\hff{\clock}{\OOt}$ between a clock and an optical oscillator is produced with a tripartite comparator (\emph{e.g.} applying the transfer oscillator technique by choosing $\rho^0_{\OOt,\clock} = m_\OOt/m_\clock$), then the transfer beat $\hfpf{\clock}{\OOt}$ for another choice $\rho'^0_{\OOt,\clock}$ of NFRs reads:
\begin{equation}
	\label{eq:ftoR}
	\hfpf{\clock}{\OOt} \simeq \hff{\clock}{\OOt} + (\rho^0_{\OOt,\clock} - \rho'^0_{\OOt,\clock})\nuz_i.
\end{equation}

\section{Overall error}
\label{sec:overallerror}

Using the approximate expression~(\ref{eq:Rapprox}) instead of the exact expression~(\ref{eq:R}), hence neglecting the impact of the (unknown) local comparator reference error $\rr_{x_i,0}$, yields an error that reads:
\begin{equation}
\delta_{\R{i-1}{i}} = \Ctri{i-1}{i}{x_i}\frac{\rho^0_{i,x_i}\rho^0_{x_i,0}}{\prod_{k=1}^i \rho^0_{k,k-1}}\rr_{x_i,0}.
\end{equation}
As stated in section~\ref{sec:comparing}, this error is practically identical to:
\begin{equation}
	\delta_{\R{i-1}{i}} \simeq \rr_{i,i-1}\rr_{x_i,0},
\end{equation}
given that $\Cshort{i-1}{i}{x_i} \simeq \rr_{i,i-1}$ and $\rho^0_{a,c} \simeq \rho^0_{a,b}\rho^0_{b,c}$ up to $10^{-6}$, which is sufficient for the error estimation.

When combining the comparator outputs to compute remote frequency ratios with~(\ref{eq:rhon0lin}), these individual error terms add-up and yield an overall error $\delta_{\rho_{n,0}}$ of the frequency ratio $\rho_{n,0}$ that is given by:
\begin{equation}
    \label{eq:error_delta}
	\frac{\delta_{\rho_{n,0}}}{\rho_{n,0}} \simeq \sum_{i=1}^n \delta_{\R{i-1}{i}}.
\end{equation}

Using the triangular inequality an upper bound for the overall error $\delta_{\rho_{n,0}}$ can be found:
\begin{equation}
	\label{eq:triangular_delta}
    \left|\frac{\delta_{\rho_{n,0}}}{\rho_{n,0}}\right| \leq \sum_{i=1}^n \left|\delta_{\R{i-1}{i}}\right|.
\end{equation}
The amount of cancellation of the individual error terms $\delta_{\R{i-1}{i}}$ in equation~(\ref{eq:error_delta}) depends on the correlation between these terms. Such correlations may lead to an overall error significantly lower than the upper bound~(\ref{eq:triangular_delta}). For instance, considering a tripartite comparator sequence between oscillators $k-1$ and $k+m$ employing physically the same RF reference $x_k$, in the oscillator-centric case the overall error is only impacted by the net output:
\begin{eqnarray*}
\sum_{i=k}^{k+m}\delta_{\R{i-1}{i}} \simeq \rr_{k+m,k-1}\rr_{x_k,0}\,.
\end{eqnarray*}
In such a comparator sequence, a misestimation of the nominal frequencies of the inner oscillators $k\ldots k+m-1$ also cancels out. This cancellation of error is also illustrated in the appendix~\ref{sec:usecase}, where we calculate the overall error for a practical example.

\section{Exact frequency ratios}
\label{sec:exactfr}

In section~\ref{sec:network}, we wrote equation~(\ref{eq:rhon0}) to express the frequency ratio $\rho_{n,0}$ as a product of local frequency ratios. This equation yields an exact frequency ratio, \emph{i.e.} without introducing an error due to the formalism. This description only fails because some comparators (in this paper, only frequency offsets and beatnotes) cannot provide exact frequency ratio measurements. The formalism presented in this paper proposes a tripartite comparator model, along with the introduction of ``accurate'' references, in order to incorporate these comparators, at the expense that the remote frequency ratios are not exact: each comparator introduces an error having the magnitude of the product of two RFRs (section~\ref{sec:comparing} and appendix~\ref{sec:overallerror}). In this section, we investigate to which extent an exact frequency ratio between oscillators can be calculated from the comparator outputs defined in our formalism, especially if the path in the network that connects these oscillators does not contain frequency beatnotes.

\subsection{Reference-free bipartite comparators}

We first assume that the path linking the oscillators $0$ and $n$ exclusively crosses reference-free bipartite comparators, \emph{i.e.} bipartite comparators that do not necessarily contain an accurate oscillator. Since the output of such comparators is none other than the reduced frequency ratio between the oscillators it compares, we can anticipate that the comparator outputs can be combined to produce the exact frequency ratio $\rho_{n,0}$. Explicitly:
\begin{eqnarray}
	\label{eq:rrexact}
	\rr_{n,0} & = & \frac{\prod_{i \in S^+}(1+\Cbi{i-1}{i})}{\prod_{i \in S^-}(1+\Cbi{i}{i-1})}-1, \\
	\textrm{where}\quad \rho^0_{n,0} & \equiv & \prod_{i=1}^n\rho^0_{i,i-1}.
\end{eqnarray}
The sets $S^+$ and $S^-$ are defined to differentiate comparators that have published outputs in the forward or backward directions with respect to the path from $0$ to $n$. A particular example is the series of two bipartite comparators linking the oscillators $a$, $b$, and $c$. In this configuration, equation~(\ref{eq:rrexact}) is written:
\begin{equation}
        \label{eq:rrexact_bipartite}
		\rr_{c,a} = \frac{\rr_{c,b} - \rr_{a,b}}{1+\rr_{a,b}}.
\end{equation}

As expected, formula~(\ref{eq:rrexact}) does not make use of the notion of accurate oscillator, that has been introduced before in order to deal with genuine tripartite comparators. Not using accurate oscillators makes the definition of $\rho^0_{n,0}$ unambiguous because the oscillators involved in the set of NFRs now exclusively have a single connection per pair (the loops in the definition of NFRs appearing in figure~\ref{fig:networkcomponents} are removed when disregarding the remote NFRs between accurate oscillators).

Because exploiting equation~(\ref{eq:rrexact}) requires arbitrary precision arithmetics, we propose another method to calculate exact frequency ratios in a chain of reference-free bipartite comparators. It is based on the relation:
\begin{equation}
	\label{eq:rrn0bipartite}
	\rr_{n,0} = \sum_{i =1}^n \Cbi{i-1}{i}\,(1+\rr_{y_i,0}),
\end{equation}
where $y_i = i-1$ if $i\in S^+$ or $y_i = i$ if $i\in S^-$. Neglecting $\rr_{y_i,0}$ is not possible because without the notion of accurate oscillators, it would lead to an error on $\rr_{n,0}$ on the order of $\epsilon^2$, where $\epsilon \simeq 10^{-6}$ is the largest linkage RFR. However, an iterative calculation for all values of $1\leq k \leq n$ starting with
\begin{equation}
	\rr^{(0)}_{k,0} = 0,
\end{equation}
and incremented by
\begin{equation}
	\label{eq:rrinduction}
	\rr_{k,0}^{(m+1)} = \sum_{i =1}^k \Cbi{i-1}{i}\,(1+\rr_{y_i,0}^{(m)}),
\end{equation}
yields an approximate value of $\rr_{n,0}$ with an error at most on the order of $\epsilon^{m+1}$ after step $m$, given that $\Cbi{i-1}{i}$ is at most on the order of $\epsilon$. Therefore, $\rr_{k,0}^{(m)}$ rapidly converges to the exact frequency ratio $\rr_{k,0}$. Equation~(\ref{eq:rrn0bipartite}) is similar to equations~(\ref{eq:rhon0lin}) and~(\ref{eq:R}). In the latter, the RFRs in the right hand side are estimated via an \emph{a priori} knowledge of the frequency ratio between designated accurate oscillators, while in the former, the RFRs are evaluated by the network itself, via a bootstrapping procedure: the evaluation of the RFRs between the oscillators of the network and the oscillator zero is iteratively refined.

\subsection{Tripartite comparators}

We now consider a path in the network that runs across a tripartite comparator. The iterative procedure does not work anymore because the reference oscillator of a tripartite comparator is not connected to the other oscillators of the network, and therefore its frequency ratio with the oscillator 0 cannot be determined solely from the comparators output. This is the reason why we introduced the notion of accurate oscillators. We can nonetheless investigate how precisely the frequency ratio $\rho_{n,0}$ can be estimated.

In section~\ref{sec:networkcomponents}, we have shown that the output of a tripartite comparator can be written as the sum of the outputs of two bipartite comparators, equation~(\ref{eq:CisCC}), if measurement capability allows to independently determine these outputs. In the examples of comparators we have shown in this paper, this only excludes frequency beatnotes and offsets, as well as frequency combs whose noise figure is too large. It can be shown from equation~(\ref{eq:R}) that the same relation holds for the quantities $\R{i-1}{i}$, whose sum yields frequency ratios over the network, assuming that the accurate reference oscillator $x_i$ of the comparator linking oscillators $i-1$ and $i$ is inserted between these two oscillators:
\begin{equation}
	\R{i-1}{i} = \R{i-1}{x_i} + \R{x_i}{i}.
\end{equation}
This relation implies that the operator of a suitable, \emph{i.e.} not genuine, tripartite comparator can declare itself as two independent bipartite comparators in series, without affecting the way frequency ratios are calculated. Using this prescription, all frequency ratios that are linked by a path that does not cross one of the genuine tripartite comparators can be exactly calculated with the standard comparator output proposed in our formalism, and the induction~(\ref{eq:rrinduction}).

If the path between the oscillators $0$ and $n$ crosses genuinely tripartite comparators, the frequency ratio $\rho_{n,0}$ can be expressed in a mixed way:
\begin{eqnarray}
	\nonumber
	\rr_{n,0} & = & \sum_{\begin{array}{c}\textrm{\small  reference-free bipartite}\\ \raisebox{0.5em}{\textrm{\small comparators}}\end{array}} \Cbi{i-1}{i}\,(1+\rr_{y_i,0}) \\ & &+
	            \sum_{\begin{array}{c}\textrm{\small genuinely tripartite}\\ \raisebox{0.5em}{\textrm{\small comparators}}\end{array}} \R{i-1}{i}.
\end{eqnarray}
These frequency ratios can be calculated iteratively, by refining the RFRs that appear in the  sum over bipartite comparators, and using the fixed approximation~(\ref{eq:Rapprox}) of $\R{i-1}{i}$ for the tripartite comparators, which still rely on the link between the accurate oscillators. The only accurate oscillators required for this are the reference oscillators of genuinely tripartite comparators and the oscillator $0$. The final frequency ratio is not exact: Tripartite comparators contribute with the same error as described in section~\ref{sec:comparing}, but now, bipartite comparators only marginally contribute to the error. This section shows that implementing bipartite comparators is preferable whenever possible because the output of a bipartite comparator does not omit information about the frequency ratios between the oscillators of the network.

\section{Use case}
\label{sec:usecase}

\begin{figure*}
\begin{center}
	\includegraphics[width=0.9\textwidth]{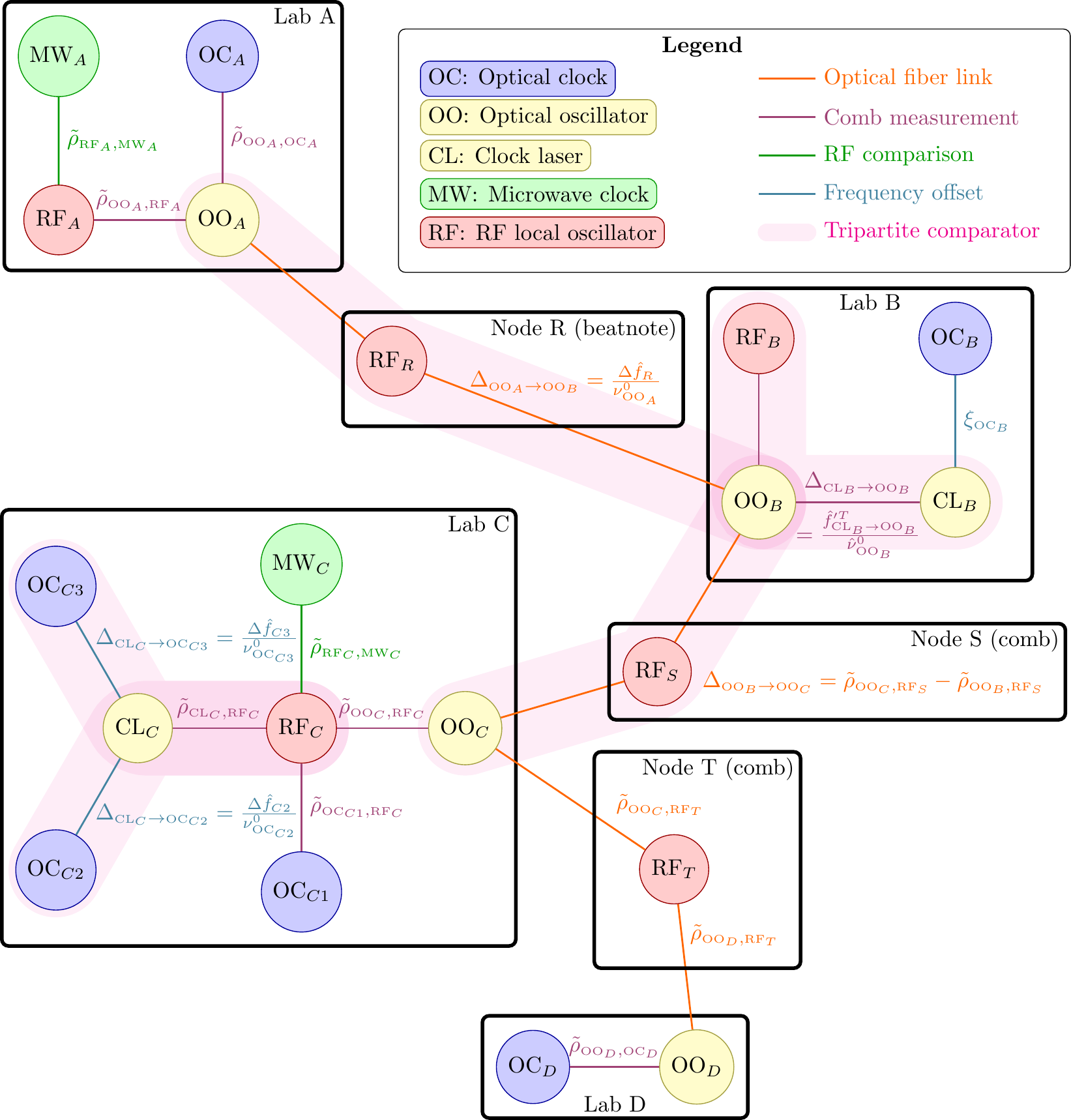}
\end{center}
\caption{\label{fig:usecase}Example architecture for a clock comparison network. The comparators are represented by lines connecting the oscillators (nodes). The ultra-stable optical oscillators (yellow nodes) are inaccurate (except $\CL_B$), while all the other oscillators are accurate, \emph{i.e} the RFRs between all these oscillators are bounded by \emph{e.g.} $\sim 10^{-13}$. The accuracy of the RF references located at the remote nodes R, S, and T is ensured by referencing them to a GPS signal. The comparator output is shown on top of each comparator line. Some comparators are tripartite, involving two optical oscillators and an RF reference, gathered by a transparent magenta area. The auxiliary reference oscillator for these tripartite comparators is the local RF oscillator. For readability, it is omitted from the notation $\C{i}{j}$ of the comparator output. The other oscillators are bipartite; their outputs are labeled with the corresponding RFR $\Cbi{i}{j} = \rr_{j,i}$ or relative systematic correction $\xi_k$. The optical oscillators are physically transported between laboratories and nodes by phase-stabilised optical fiber links.}
\end{figure*}

We finally illustrate how to effectively implement the formalism set out in this paper with a fictitious comparison network that is used to compare four laboratories, $A$, $B$, $C$, and $D$, as depicted in figure~\ref{fig:usecase}.

The various laboratories are equipped with optical and microwave clocks and oscillators, connected by frequency combs (purple lines). The optical oscillators are typically lasers stabilized on ultra-stable cavities. Some of them -- the ``clock lasers'' -- are designed such that their frequency or a sub-harmonic lies nearby the resonance frequency of an optical clock. At $B$, the clock laser $\CL_B$ is locked to the optical clock $\OC_B$, before the relative systematic correction $\xi_{\OCt_B}$, on the order of $10^{-15}$ has been applied. At $C$, the same clock laser $\CL_C$ is shared by two clocks $\OC_{C2}$ and $\OC_{C3}$ using the same atomic species. The RF offsets $\Delta \hat f_{C2}$ and $\Delta \hat f_{C3}$ are applied to bridge the gap between the common clock laser and each optical clock. We assume that the systematic correction of the other clocks in the network are locally incorporated in the comparator output, and therefore not exposed in the data sharing. In the four laboratories, all comparators are bipartite except for the frequency offsets mentioned above that must be referenced to the local RF oscillator $\RF_C$, and the comparison between the clock laser $\CL_B$ and the optical oscillator $\OO_B$ for which the comparator is chosen tripartite, referenced to $\RF_B$.

The optical oscillators are transported by phase-stabilized optical fiber links (orange lines), and are connected at three nodes $R$, $S$, and $T$. The frequency difference between $\OO_A$ and $\OO_B$ is small enough to be counted with a frequency counter at node $R$, whose output is noted $\Delta \hat f_R$. The oscillators $\OO_B$ and $\OO_C$ are operated on different telecom channels, and bridged by a frequency comb at node $S$, operated as a tripartite comparator. The node $T$ compares the oscillators $\OO_C$ and $\OO_D$ with a frequency comb operated as a series of two bipartite comparators.

In order to write the comparisons between the oscillators in a concise form, we assume that network-wide nominal frequencies have been defined by the network collaboration prior to operation, as described in section~\ref{sec:oscillatorcentric}. However, a similar treatment could be conducted with independently chosen NFRs, and by computing the frequency ratios using the general formulas of section~\ref{sec:computingratios}. In order to quantify the errors introduced by the formalism, we propose an example for the frequency map for the various oscillators. The frequencies of the optical oscillators would be chosen close to telecom channels in the C band, such that the collaboration can choose the nominal frequencies:
\begin{eqnarray}
	\nuz_{\OOt_A} = \nuz_{\OOt_B} = &\ 194.4\ex{12}, \\
	\nuz_{\OOt_C} = \nuz_{\OOt_D} = &\ 194.6\ex{12}.
\end{eqnarray}
For an oscillator $X$, we note $\epsilon_X$ the order of magnitude of $\rr_{X,Z}$, where $Z$ is an arbitrary clock. The typical frequency offsets from the fiber channels would then be, for instance, in the unit $u_Z$:
\begin{eqnarray}
	\nuz_{\OOt_A}\epsilon_{\OOt_A} \simeq -45\ex{6}, & \quad
	\nuz_{\OOt_B}\epsilon_{\OOt_B} \simeq -40\ex{6}, \\
	\nuz_{\OOt_C}\epsilon_{\OOt_C} \simeq  -2\ex{6}, & \quad
	\nuz_{\OOt_D}\epsilon_{\OOt_D} \simeq  20\ex{6}.
\end{eqnarray}
\emph{i.e.}
\begin{eqnarray}
	\epsilon_{\OOt_A} \simeq -2.3\ex{-7}, \quad &
	\epsilon_{\OOt_B} \simeq -2.1\ex{-7}, \\
	\epsilon_{\OOt_C} \simeq -10^{-8}, \quad &
	\epsilon_{\OOt_D} \simeq 10^{-7}.
\end{eqnarray}
The frequencies of $\OO_A$ and $\OO_B$ are chosen close enough to relax the required accuracy of the calibration of the RF reference $\RF_R$: its inaccuracy is rejected by $\epsilon_{\OOt_B} - \epsilon_{\OOt_A} \simeq 3\ex{-8}$.

The nominal frequency of a given clock laser is set to the nominal frequencies of its respective clock. In particular, the nominal frequencies for the clocks of laboratory $C$ are chosen such that $\nuz_{\OCt_{C2}} = \nuz_{\OCt_{C3}} = \nuz_{\CLt_{C}}$. The comparator outputs $\C{\CLt_C}{\OCt_{C2}}$ and $\C{\CLt_C}{\OCt_{C3}}$ can be on the order of $10^{-6}$.

\subsection{Data acquisition and publication}

The operator of each subsystem ensures before a measurement campaign or on a regular basis that her or his counters acquire the data synchronized to a 1~pps signal, which is derived e.g. from UTC(k) or from GPS, such that it is sufficiently synchronized to the counters used by the other subsystems. Furthermore, all counters must be operated in the same averaging mode (e.g. $\Pi$ or $\Lambda$~\cite{ben15}), which must be agreed upon beforehand with the other operators. The data are published as datasets with the comparator output on a time-stamped, previously agreed UTC 1~pps -- synchronous grid (e.g. 1~s or 10~s). Optionally, the time stamped data can contain a validity flag marking the data as valid or invalid, e.g. due to downtimes of the respective subsystem.

The data sets are produced as follows:
\begin{itemize}
	\item Alicia operates comb $A$, comparing the optical oscillator $\OO_A$ to the optical clock $\OC_A$ and to the RF reference $\RF_A$. She produces and publishes the reduced ratios $\rr_{\OOt_A,\RFt_A} = \hat\nu_A/\nuz_A - 1$ using equation~(\ref{eq:hatnucomb}), and $\rr_{\OOt_A,\OCt_A}$ following~(\ref{eq:rrexact_bipartite}) from
	\begin{equation}
		\label{eq:alicia}
		\rr_{\OOt_A,\OCt_A} = \frac{\rr_{\OOt_A,\RFt_A} - \rr_{\OCt_A,\RFt_A}}{1+\rr_{\OCt_A,\RFt_A}}.
	\end{equation}
	As a verfication, she multiplies each quantity by $\nuz_{\OOt_A} = 194.4\ex{12}$ and finds a transfer beat around $-45\ex{6}$, close to the documented frequency offset of $\OO_A$, as expected.
	\item Albus operates the microwave clock $\MC_A$. He produces $\rr_{\RFt_A, \MCt_A}$ as he usually does, shares the data with the link collaboration, and sends the average to the BIPM as a calibration of TAI.
	\item Bathilda operates comb $B$. She estimates that for her comb, the transfer oscillator computation is advantageous. She produced the transfer beat $\hff{\CLt_B}{\OOt_B}$ according to~(\ref{eq:fftransosc}), and transforms it to with~(\ref{eq:ftoR}) to comply with the collaboration's choice of NFRs based on nominal frequencies. She then computes the comparator output $\C{\CLt_B}{\OOt_B} = \hfpf{\CLt_B}{\OOt_B}/\nuz_{\OOt_B}$, and publishes the dataset.
	\item Bill evaluates the systematic effects of clock $\OC_B$ and publishes the relative frequency correction $\xi_{\OCt_B}$ to be applied to the clock laser $\CL_B$.
	\item Unlike Alicia who generates the RF-free optical-to-optical frequency ratio $\rr_{\OOt_A,\OCt_A}$ with equation~(\ref{eq:alicia}), Cedric, the operator of the comb $C$, prefers to use the linearised equation $\rr_{\OOt_C,\OCt_C} \simeq \rr_{\OOt_C,\RFt_C} - \rr_{\OCt_C,\RFt_C}$ to compute frequency ratios because he estimates that the linearisation error is negligible in his system: his RF reference is well calibrated and the frequency of $\OO_C$ is close to its nominal value. He therefore calculates the reduced frequency ratios $\rr_{\OOt_C, \RFt_C}$, $\rr_{\OCt_{C1}, \RFt_C}$, and $\rr_{\CLt_C, \RFt_C}$, and publishes them. He also sends the average of $\rr_{\OCt_{C1}, \RFt_C}$ to the BIPM to calibrate TAI with the optical clock $\OC_{C1}$.
	\item Charlie operates the optical clocks $\OC_{C2}$ and $\OC_{C3}$ by applying frequency corrections to a pair of acousto-optic modulators. He \emph{a posteriori} adds to these frequencies the corrections for the clocks' systematic effects, and rescales the outcome by the nominal clock frequency $\nuz_{\OCt_{C2}} = \nuz_{\OCt_{C3}}$ to obtain $\C{\CLt_C}{\OCt_{C2}}$ and $\C{\CLt_C}{\OCt_{C3}}$, and publishes them.
	\item Draco is in charge of the optical clock $\OC_D$. It is a composite clock that probes two different clock transitions in order to cancel the sensitivity to the black-body radiation. Together with Dolores, the operator of comb $D$, they designed an astute scheme involving a combination of comb beatnotes to synthesize the clock's frequency. With it, they are able to produce and publish $\rr_{\OOt_D,\OCt_D}$. Albus, who is not familiar with optical frequencies, is relieved that he will not have to understand in detail how a synthetic optical clock works to calculate the absolute frequency of $\OC_D$ with his Cs clock $\MC_A$.
	\item Romilda is in charge of node $R$. She collects the counter output, rescales it by the nominal frequency $\nuz_A = \nuz_B$, and publishes it as $\C{\OOt_A}{\OOt_B}$.
	\item Sybill operates the comb of node $S$ used as a genuine tripartite comparator. She produces the comparator output $\C{\OOt_B}{\OOt_C} = \rr_{\OOt_C, \RFt_S} - \rr_{\OOt_B, \RFt_S}$ and publishes it.
	\item Unlike Sybill, Tom exposes his frequency comb comparing $\OO_C$ and $\OO_D$ at node T as two bipartite comparators. He individually measures $\rr_{\OOt_C, \RFt_T}$ and $\rr_{\OOt_D, \RFt_T}$, and publishes both of them.
\end{itemize}

\subsection{Comparing oscillators}
We can now express the reduced frequency ratios between various oscillators, using the relation~(\ref{eq:RC}). For each case, we specify the error due to the use of reduced frequency ratios.
\begin{itemize}
	\item Comparing $\OC_B$ to $\OC_A$ amounts to calculating:
	\begin{eqnarray}
		\nonumber
		\rr_{\OCt_B,\OCt_A} \simeq &&\ \rr_{\OOt_A,\OCt_A} + \C{\OOt_A}{\OOt_B} \\ && + \C{\OOt_B}{\CLt_B} + \xi_{\OCt_B}.
	\end{eqnarray}
	The errors on these terms are on the order of 0, $(\epsilon_{\OOt_B} - \epsilon_{\OOt_A})\epsilon_{\RFt_R}$, $\epsilon_{\OOt_B}\ec$, and $\ec^2$, respectively. We recall that $\ec \sim 10^{-15}$ is the order of magnitude of the reduced frequency ratios between clocks. Note that because Bathilda uses the transfer oscillator method to produce $\C{\OOt_B}{\CLt_B}$, this term contains an additional error of $\epsilon_b\epsilon_{\RFt_B}$, where $\epsilon_b \sim 10^{-6}$.
	\item Comparing $\OC_D$ to $\MC_A$ is achieved by:
	\begin{eqnarray}
		\nonumber\rr_{\OCt_D,\MCt_A} \simeq & &\  \rr_{\RFt_A,\MCt_A} + \rr_{\OOt_A,\RFt_A} \\
		\nonumber&&\label{eq:rrODCMCA}
		 + \C{\OOt_A}{\OOt_B} + \C{\OOt_B}{\OOt_C}  \\ && + \rr_{\OOt_D,\RFt_T} - \rr_{\OOt_C,\RFt_T} - \rr_{\OOt_D,\OCt_D}.
	\end{eqnarray}
	The errors are 0, $\epsilon_{\OOt_A}\epsilon_{\RFt_A}$, $(\epsilon_{\OOt_B} - \epsilon_{\OOt_A})\epsilon_{\RFt_R}$, $(\epsilon_{\OOt_C} - \epsilon_{\OOt_B})\epsilon_{\RFt_S}$, $(\epsilon_{\OOt_D} - \epsilon_{\OOt_C})\epsilon_{\RFt_T}$, and $\epsilon_{\OOt_D}\ec$, respectively
	\item Comparing $\OC_{C3}$ to $\OC_{C2}$:
	\begin{equation}
		\rr_{\OCt_{C3},\OCt_{C2}} \simeq \C{\CLt_C}{\OCt_{C3}} - \C{\CLt_C}{\OCt_{C2}}.
	\end{equation}
	The error is on the order of $\ec\epsilon_{\RFt_C}$, which is completely negligible.
	\item Comparing $\OC_{C1}$ to $\OC_{C2}$:
	\begin{eqnarray}
		\nonumber
		\rr_{\OCt_{C1},\OCt_{C2}} \simeq &&\  -\C{\CLt_C}{\OCt_{C2}} - \rr_{\CLt_{C},\RFt_C} \\&& + \rr_{\OCt_{C1},\RFt_C}.
	\end{eqnarray}
	The error is again on the order of $\ec\epsilon_{\RFt_C}$, given that all comparators are referenced to the same RF source (see appendix~\ref{sec:overallerror}). This comparison does not involve the $\OO_C$ optical oscillator.
	\item Comparing $\OC_{B}$ to $\OC_{C2}$:
	\begin{eqnarray}
		\label{eq:overallerrorusecase}
		\nonumber \rr_{\OCt_{B},\OCt_{C2}} \simeq &&\ -\C{\CLt_C}{\OCt_{C2}} - \rr_{\CLt_{C},\RFt_C} \\ \nonumber &&+ \rr_{\OOt_C,\RFt_C}  - \C{\OOt_B}{\OOt_C} \\&& + \C{\OOt_B}{\CLt_B} + \xi_{\OCt_B}.
	\end{eqnarray}
	The error of the last three terms is the same as above. For the first three terms, we can here again consider their overall error (see appendix~\ref{sec:overallerror}), which can be calculated by noticing that these terms can be grouped as:
	\begin{eqnarray}
		\nonumber
		\rr_{\OOt_C,\RFt_C} - \rr_{\CLt_{C},\RFt_C} - \C{\CLt_C}{\OCt_{C2}} &&\\ \label{eq:CLOCC2}= \rr_{\OOt_C,\OCt_{C2}}(1+\rr_{\OCt_{C2},\RFt_C}),&&
	\end{eqnarray}
	which differs from the desired $\rr_{\OOt_C,\OCt_{C2}}$ by an error $\epsilon_{\OOt_C}\epsilon_{\RFt_C}$.
	\item Comparing $\MC_C$ to $\MC_{A}$:
	\begin{eqnarray}
		\nonumber \rr_{\MCt_C,\MCt_A} \simeq &&\  \rr_{\RFt_A,\MCt_A} + \rr_{\OOt_A,\RFt_A}  \\ \nonumber && + \C{\OOt_A}{\OOt_B} + \C{\OOt_B}{\OOt_C} \\ && - \rr_{\OOt_C,\RFt_C} - \rr_{\RFt_C,\MCt_C}.
	\end{eqnarray}
	\item Comparing $\OO_D$ to $\OO_A$ is a bit more involving, because the oscillator $\OO_A$ is not an accurate oscillator such that the approximation~(\ref{eq:Rapprox}) does not hold. To go around this issue, one can write:
	\begin{equation}
		\rr_{\OOt_D,\OOt_A} = \frac{\rr_{\OOt_D, Z} - \rr_{\OOt_A, Z}}{1 + \rr_{\OOt_A, Z}},
	\end{equation}
	where $Z$ is a hypothetical clock connected to $\OO_A$. The numerator can be computed as before, while the denominator can simply be approximated to 1 because the comparison of optical oscillators is not meant to be accurate. Finally:
	\begin{eqnarray}
		\nonumber
		\rr_{\OOt_D, \OOt_A} \simeq &&\ \C{\OOt_A}{\OOt_B} + \C{\OOt_B}{\OOt_C} \\ && + \rr_{\OOt_D,\RFt_T} - \rr_{\OOt_C,\RFt_T}.
	\end{eqnarray}
	\item OO$_D$ can be compared to OC$_{C1}$ without introducing an error due to the formalism because they are solely connected by bipartite comparators. This can be done with the iterative procedure described in appendix~\ref{sec:exactfr}, or with equation~(\ref{eq:rrexact}):
	\begin{equation}
		\rr_{\OOt_D,\OCt_{C1}} = \frac{(1+\rr_{\OOt_C,\RFt_C})(1+\rr_{\OOt_D,\RFt_T})}{(1+\rr_{\OCt_{C1},\RFt_C})(1+\rr_{\OOt_C,\RFt_T})} - 1.
	\end{equation}
	\item even though the path linking OO$_D$ to OC$_{C2}$ contains a genuine tripartite comparator (the frequency offset between OC$_{C2}$ and CL$_C$ referenced to RF$_C$), their frequency ratio can nonetheless be derived without introducing a formalism error because the frequency ratio between CL$_C$ and RF$_C$ is independently measured by a bipartite comparator. Using equation~(\ref{eq:rrexact}) yields:
	\begin{eqnarray}
		\nonumber
		\rr_{\OOt_D,\OCt_{C2}} = &&\ \frac{1}{1+\rr_{\OOt_C,\RFt_T}}(\rr_{\OOt_C,\OCt_{C2}} \\ \nonumber && +\rr_{\OOt_D,\RFt_T} - \rr_{\OOt_C,\RFt_T} \\ &&+ \rr_{\OOt_C,\OCt_{C2}}\rr_{\OOt_D,\RFt_T}),
	\end{eqnarray}
	where $\rr_{\OOt_C,\OCt_{C2}}$ is given by equation~(\ref{eq:CLOCC2}).
\end{itemize}

The examples above illustrate the statement that the significant errors introduced by the formalism are proportional to the mismatch between the optical oscillators and their nominal frequencies. Reducing the error thus amounts to changing the nominal frequencies to better match the actual frequencies of the optical oscillators. This recentering is limited by the magnitude of the beat frequencies in the nodes of the network.

Finally, the gravitational redshift can be taken into account by adding:
\begin{equation}
	\Delta y_\textrm{\tiny GRS} = \frac{C_i - C_j}{c^2}
\end{equation}
to the reduced frequency ratios $\rr_{j,i}$, where $C_i$ is the geopotential number at the location of clock $i$, expressed in a common reference frame~\cite{Delva2019}.

\bibliography{freqratios}

\end{document}